\documentclass[aps,pre,amsmath,amssymb,nofootinbib,superscriptaddress,twocolumn]{revtex4-1}
\usepackage{graphicx}
\usepackage{amsmath,pgfmath,pgffor}
\usepackage[usenames, dvipsnames]{color}
\usepackage{indentfirst}
\usepackage{hyperref}
\usepackage{subfigure}
\usepackage{epstopdf}

\usepackage{hyperref}

\newcommand{\be}{\begin{equation}}
\newcommand{\ee}{\end{equation}}

\usepackage{color}

\hypersetup{%
    pdfborder = {0 0 0},
    colorlinks,
    citecolor=blue,
    filecolor=Darkgreen,
    linkcolor=black,
}

 \let\b=\beta \let\g=\gamma 
  \let\h=\eta \let\k=\kappa
\let\l=\lambda    
\let\s=\sigma  \let\f=\varphi 
   
\let\D=\Delta   
   
\let\ee=\varepsilon  \let\th=\theta \let\io=\infty

\def\to{\rightarrow}

\newcommand{\beq}{\begin{equation}} \newcommand{\eeq}{\end{equation}}
\newcommand{\wh}{\widehat}

\begin{document}

\title{Mean field stability map of hard sphere glasses}

\author{Ada Altieri}
\affiliation{Laboratoire de Physique de l'\'Ecole Normale Sup\'erieure, Universit\'e PSL, CNRS, Sorbonne Universit\'e, Universit\'e de Paris, 75005 Paris, France}
\author{Francesco Zamponi}
\affiliation{Laboratoire de Physique de l'\'Ecole Normale Sup\'erieure, Universit\'e PSL, CNRS, Sorbonne Universit\'e, Universit\'e de Paris, 75005 Paris, France}

\begin{abstract}

The response of
amorphous solids to an applied shear deformation is an important problem, both in fundamental and applied research. 
To tackle this problem, we focus on a system of hard spheres in infinite dimensions as a solvable model for colloidal systems and granular matter. 
The system is prepared above the dynamical glass transition density, and we discuss the phase diagram of the resulting glass 
under compression, decompression, and shear strain, expanding on previous results
[P.Urbani and F.Zamponi, Phys.Rev.Lett. {\bf 118}, 038001 (2017)].
We show that the solid region is bounded by a ``shear jamming'' line, at which the solid reaches close packing, 
and a ``shear yielding'' line, at which the solid undergoes a spinodal instability towards a liquid, flowing phase. Furthermore, we characterize
the evolution of these lines upon varying the glass preparation density.
This work aims to provide a general overview on yielding and jamming phenomena in hard sphere systems by a systematic exploration of the phase diagram.
\end{abstract}

\maketitle

\section{Introduction} 
A glass can be essentially described as an amorphous solid characterized by an enormous slowing down of the diffusive motion. This slowing down is accompanied by a spectacular increase of the relaxation time and the viscosity. 
The glass transition is empirically defined as the temperature below which (or density above which) the material becomes 
too viscous to undergo a diffusive motion on experimentally relevant timescales, and is then frozen in an out-of-equilibrium state.

Because the glass transition is not a sharp phase transition, but rather an anthropocentric concept, and the glass is an out-of-equilibrium state,
protocol preparation plays a strong role in determining glass properties. 
As a result, glassy materials display peculiar features distinct from ordinary solids. 
For instance, the stress-strain curves of the glass strongly depend on its degree of annealing~\cite{RTV11,BDBDM17,OBBRT18}.
Furthermore, the response of glasses to an external weak stress is characterized by large deformations and very complex phenomena that can be generally explained by invoking marginal stability~\cite{LN10,MW15}. The marginal stability condition is responsible for unusual properties ranging from an 
abundance of zero-frequency modes~\cite{WNW05} to the emergence of long-range correlations in vibrational dynamics~\cite{BCJPSZ16}. 

In the following, we will focus on a thermodynamic approach to structural glasses, originally proposed and developed by 
Kirkpatrick, Thirumalai and Wolynes~\cite{KW87, KT87, KTW89}, and known as the Random First Order Transition (RFOT) theory.
Within this theory, glasses are considered as extremely long-lived metastable states, 
essentially described in terms of vibrations around an amorphous
reference structure~\cite{SSW85}, which encodes the protocol dependence.
Because here we consider glasses obtained by slow cooling of a supercooled liquid, the reference structure
is taken to be a typical equilibrium configuration at the state point where the liquid fell out-of-equilibrium~\cite{RUYZ15}; such a construction is
called the Franz-Parisi potential method, originally developed in spin glass theory~\cite{FP95, FP97, BFP97, KZ10, KZ10b, KZ13}.

In the abstract limit of infinite spatial dimension $d$, the method is exact~\cite{CKPUZ17}, but it can also be used in two or three dimensions in an approximate
way~\cite{MP96,CFP98,YM10,PZ10}.
Results for infinite-dimensional hard spheres~\cite{RUYZ15,UZ17} show
that the application of a strain $\gamma$ can result in very different and highly complex scenarios.
Generically, at small~$\gamma$, an elastic-like regime is observed, which is identified by a linear dependence between the response stress $\sigma$ and the applied strain, i.e. $\sigma \sim \mu \gamma$ with $\mu$ being the shear modulus~\cite{YZ14}.
Upon increasing the strain, beyond a Gardner transition~\cite{Ga85,KPUZ13,nature,RUYZ15,RU16,BU16,BU18},
the response turns out to be characterized by intermittent drops or avalanches, related to
collective rearrangements of the whole system,
which is nevertheless still solid~\cite{RU16,FS17}. Depending on the preparation and target densities, two scenarios are possible at larger strain.
Either the system eventually looses its solidity and starts to flow, undergoing \emph{shear yielding}, or it can reach a jamming (or close packing) point,
where a rigid network of contacts is formed and the system can no longer be strained~\cite{UZ17}. 
The shear jamming and the isotropic jamming (under compression without shear) transitions fall in the same universality class 
with the same critical exponents regulating the distributions of effective forces and gaps~\cite{UZ17}. 
Furthermore, the theory has been extended to hard sphere with a short range attraction~\cite{SZ13},
to investigate a peculiar two-step yielding transition that characterizes colloidal systems~\cite{AUZ18}. 
These predictions, obtained in the mean-field $d\to\io$ limit, have been partially confirmed by extensive numerical simulations in $d=3$~\cite{CCPZ12,CCPZ15,BGLNS16,JY17,JUZY18}
and experiments on a granular material in $d=2$~\cite{SD16}.

In this paper, we will complete and extend the hard-sphere results obtained in~\cite{RUYZ15,UZ17}. We will explore a wider regime of densities, in order to obtain 
a complete phase diagram. In particular, we will show that the shear yielding line terminates on the melting point at low density and vanishing shear strain, 
while on the other side it touches the shear jamming line 
at a critical point identified in~\cite{UZ17}. We will study the preparation density evolution of these lines, 
and we will provide further details on the nature of the Gardner transition already studied in~\cite{RU16}.

The article is organized as follows: in Sec.~\ref{SecII}, we introduce the mathematical formalism for a system of hard spheres in $d$ dimensions, with $d\to\io$; in Sec.~\ref{SecIII}, 
we describe the relevant phase transitions, including the jamming, yielding and Gardner transitions. Then, in Sec.~\ref{SecIV} we present our results, based on the determination of the glass phase diagrams, under compression or decompression and quasi-static shear strain. 
In Sec.~\ref{sec:conc} we conclude presenting some perspectives for future investigations.

\section{State following procedure}
\label{SecII} 

We consider a system of $N$ identical hard spheres in $d$ dimensions, which turns out to be a very convenient model for structural glasses and granular matter
(monodisperse spheres are a good glass former as long as $d\geq 4$~\cite{SDST06,VCFC09}). 
We then consider the limit $d \rightarrow \infty$, which makes the problem analytically tractable and 
provides an exact thermodynamical~\cite{KPZ12, RUYZ15} and dynamical solution~\cite{MKZ16,AMZ19}.
Note that, in order to obtain a finite $d\to\io$ limit, one has to rescale the packing fraction $\f$, which is the fraction of volume occupied by the spheres, introducing
 $\widehat{\varphi}=2^{d} \varphi/d$. The other observables have to be rescaled as well, as detailed below. 
 For hard spheres, thermal fluctuations are irrelevant, and the only control parameter in equilibrium is density. 

To perform our analysis we employ a \emph{state following} protocol~\cite{FP95, FP97, BFP97, KZ10, KZ10b, KZ13}, 
which describes glass formation during a very slow liquid cooling (see~\cite{RUYZ15} for a detailed explanation in the context of structural glasses). 
The state following method works for very large relaxation times, which is actually the case in infinite dimensions where the equilibrium relaxation time diverges upon approaching the dynamical transition at packing fraction~$\wh\f_{\rm d}$~\cite{MKZ16,CKPUZ17} (or critical transition in Mode-Coupling Theory~\cite{Go99}). 
Several properties of glassy systems can be re-obtained in this framework, such as the stress overshoot and the shear modulus behaviors, as well as the dependence of the pressure and the specific volume on the cooling rate~\cite{RUYZ15,CKPUZ17}.

Because the method has been discussed in details elsewhere~\cite{RUYZ15,CKPUZ17}, we only provide here a brief summary of the main equations.
We consider a system equilibrated at a value of density $\widehat{\varphi}_g > \widehat{\varphi}_{\rm d}$, where diffusion 
 is frozen and an enormous number of metastable glassy states emerge,
 each with the same pressure as the equilibrium liquid prepared at $\widehat{\varphi}_{\rm d}$. 
  In this region, the system is trapped in one such metastable state, 
whose evolution under compression/decompression and/or an applied shear strain is then studied~\cite{RUYZ15}.

To realize this construction in practice, we first extract an equilibrium configuration, $Y =\{\bf{y}_i\}$, from the Boltzmann-Gibbs measure at $\widehat{\varphi}_g$, and then we consider another configuration, $X=\{\bf{x}_i\}$, which is constrained to be close to the former. Their mutual mean-squared displacement is fixed by the condition $\Delta(X, Y)= \Delta_r$, where 
$\Delta (X, Y) = \frac{d}{N} \sum \limits_{i=1}^{N} \vert {\bf{x}}_i-{\bf{y}}_i \vert ^2$.
The resulting partition function then becomes:
\begin{equation}
Z[\Delta_r, \widehat{\varphi},\g \vert Y, \widehat{\varphi}_g]= \int d X e^{-\beta V_\g[X; \widehat{\varphi}]} \delta  \left( \Delta_r - \Delta(X,Y) \right) \ ,
\label{Z_g}
\end{equation}
where $V_{\g=0}[X;\wh\f]$ is a hard-core potential, at a density $\wh\f$ that can differ from $\wh\f_g$, thus realizing a compression or decompression.
An applied shear strain $\g$ on configuration $X$ is described by a modified potential $V_\g[X;\wh\f]$, in which a linear transformation is applied to configuration $X$
to describe the box straining, 
as derived in~\cite{YM10,RUYZ15}. 
In this way, an equilibrium theory for the rheology of amorphous solids is obtained, by keeping the reference configuration $Y$ unstrained and applying 
a strain on the other, constrained replica $X$. The physical idea behind this construction is that in the solid phase, after the application of the affine shear strain,
the system relaxes via non-affine deformation to a stationary state which is close enough to the original state; the constrained replica $X$ describes precisely
this long-time pseudo-equilibrium state. See~\cite{YM10,Yo12} for a detailed discussion.

Unfortunately, the computation of the partition function in Eq.~(\ref{Z_g}), as well as that of the associated free energy, 
is impossible because the constraint $\D_r = \D(X,Y)$, for fixed $Y$, explicitly breaks the translational invariance of the measure for $X$:
the equilibrium configuration $Y$ plays the role of a quenched disorder. Yet, the free energy is self-averaging with respect to $Y$, which means that
for given $\wh\f_g$ and $ (\wh\f,\g)$, the free energy for given $Y$ is equal to that averaged over $Y$ with probability one in the thermodynamic limit.
One can then take the equilibrium average over $Y$, which can be done via
the replica method~\cite{FP95, MPV87}. This procedure restores translational invariance and makes the problem tractable~\cite{RUYZ15}.
One then introduces $s$ replicas of configuration $X$ and eventually considers the analytic continuation to $s \rightarrow 0$. 
In order to perform this analytic continuation, an ansatz has to be made on the behavior of the replicas under the permutation symmetry~\cite{MPV87},
the simplest case being the so-called replica symmetric (RS) ansatz, in which the replicas are symmetric under any permutation~\cite{MPV87}.
A single order parameter, $\D = \D(X_a,X_b)$, has to be introduced to describe the mean-squared displacement between any pair $a,b$ of replicas of $X$.
Once the analytic continuation is performed, the averaged free energy should be extremized with respect to the variational parameters $\D,\D_r$. 
Additional technical details can be found in~\cite{RUYZ15, RU16}.

In the hard-sphere model considered thus far,
energy vanishes and only entropy is relevant.
The RS glass entropy (per particle and per spatial dimension, i.e. per degree of freedom) 
can be exactly computed in the limit $d\to\io$, and is given by
\begin{equation}
\begin{split}
s_g &=   \frac{\Delta_r-\Delta/2}{\Delta} + \frac{1}{2} \log \left(\frac{\pi e \Delta}{d^2} \right) + \\
& + \frac{  \widehat{\varphi}_g}{2} \int_{-\infty}^{\infty} dh e^{h} q_{\gamma} \left( 2 \Delta_r-\Delta; h \right) f(h) \ .
\end{split}
\label{entropy}
\end{equation}
The function $q(\Delta; h)$  reads for $\g=0$
\begin{equation}
q(\Delta; h)= \Theta \left( \frac{h +\Delta/2 }{\sqrt{ 2 \Delta}} \right) \ ,
\label{q}
\end{equation}
with
$\Theta(x)=  \frac{1}{2}\left( 1+ \text{erf}(x) \right)$,
and the shorthand
notation $f(h)= \log q \left( \Delta; h -\eta \right)$ is used in Eq.~\eqref{entropy} and in the following. The parameter $\eta=\log(\widehat{\varphi}/\widehat{\varphi}_g)$ denotes the logarithm of the relative compression (if positive) or decompression (if negative).
The function $q_{\gamma}(\Delta; h)$, which encodes the dependence on 
a shear strain $\gamma$, is given by
\begin{equation}
q_{\gamma}(\Delta; h) = \int \mathcal{D} \zeta q( \Delta + \gamma^2 \zeta^2; h) \ ,
\end{equation}
where $\mathcal{D} \zeta$ denotes the Gaussian measure with zero mean and unit variance.

The differentiation of Eq. (\ref{entropy}) with respect to the self mean-square displacement, $\Delta$, and the relative mean-square displacement, $\Delta_r$, provides 
two equations
\begin{equation}
\begin{split}
2 \Delta_r  =& \Delta + \widehat{\varphi}_g \Delta^2 \times \\
& \times \int_{-\infty}^{\infty} dh e^{h} \frac{\partial}{\partial \Delta} \left[ q_{\gamma}(2 \Delta_r- \Delta; h ) 
 f(h) \right] \ , \\
\frac{2}{\Delta}  =& -\widehat{\varphi}_g \int_{-\infty}^{\infty} dh e^{h} \left[ \frac{\partial}{\partial \Delta_r} q_\gamma \left( 2\Delta_r -\Delta; h \right)  \right] f(h)  \ ,
\label{iterative-eqs}
\end{split}
\end{equation}
which can be solved iteratively to determine $\D,\D_r$.
Because the existence of a unique solution is not guaranteed (see~\cite{AUZ18} for an explicit example), 
we start from the equilibrium solution and follow it adiabatically upon increasing $\gamma$. 

At equilibrium $(\wh\f=\wh\f_g, \g=0)$, we have $\Delta=\Delta_r$,  and Eqs.(\ref{iterative-eqs}) reduce to a single 
equation
\begin{equation}\label{eq:equi}
\frac{1}{\widehat{\varphi}}= -\Delta \int_{-\infty}^{\infty} dh e^{h} \log \left[ q( \Delta; h) \right]  \frac{\partial q(\Delta; h)}{\partial \Delta} \equiv \mathcal{F}(\Delta) \ .
\end{equation}
The function $\mathcal{F}(\Delta)\geq 0$ vanishes at $\Delta=0$ and $\Delta \rightarrow \infty$, and it therefore has a maximum in between, 
whole value provides the dynamical transition density $1/\widehat{\varphi}_{\rm d}=4.8067$:
\begin{equation}
\frac{1}{\widehat{\varphi}_{\rm d}} = \max_\Delta \mathcal{F}(\Delta) \ .
\label{dyn}
\end{equation}
For $\wh\f<\wh\f_{\rm d}$, Eq.~(\ref{dyn}) has no solution, which translates into the impossibility of finding a stable glass phase: the system is liquid.
Instead, for $\wh\f>\wh\f_{\rm d}$, a solution for $\D$ exists in equilibrium. We then need to select the initial glass state at $\widehat{\varphi}_g>\widehat{\varphi}_{\rm d}$. 

To iterate Eqs.(\ref{iterative-eqs}) within a given precision for $\wh\f\neq \wh\f_g$ and/or $\g>0$, 
we set as initial condition $\D=\D_r$ equal to the solution of Eq.~\eqref{eq:equi} at $(\wh\f=\wh\f_g, \g=0)$,
we compute numerically the right hand side of Eqs.(\ref{iterative-eqs}) at the new state point, and we iterate until convergence.
Then we slowly change the density, the shear strain or both, and we use the values obtained at the previous step as new guess. 
Once we have checked the convergence of the above equations and obtained the resulting values of $\Delta$ and $\Delta_r$, we can compute 
the glass entropy, its reduced pressure per degree of freedom $\wh p_g = \b P_g/(d \rho)$, 
\begin{equation}
\wh p_g=-\frac{ \widehat{\varphi}_g}{2} \int_{-\infty}^{\infty} dh e^{h} q_{\gamma} \left( 2 \Delta_r-\Delta; h \right) \frac{\partial f(h) }{\partial \eta}
\end{equation}
and all other thermodynamic quantities of interest.

\section{Phase transitions}
\label{SecIII}

We now aim at completely characterizing the phase diagram of a glass prepared at $\wh\f_g$, and followed in compression/decompression and
applied shear strain, $(\wh\f,\g)$.
 In the following, we briefly recap the main features of the phase transitions that are encountered in this hard-sphere system.

\subsection{Jamming}
 \label{sec:Jdef}

Conventionally, the jamming transition line of hard spheres is identified as the locus of points where the pressure diverges 
and the mean-square displacement approaches zero~\cite{PZ10}. 
We can then define a critical shear jamming line $\gamma_{J}(\eta)$, or $\eta_J(\g)$, in correspondence of which the jamming transition takes place. 
We take the limit $\Delta \rightarrow 0$ in Eqs.~\eqref{iterative-eqs} and, using the relation
\beq
\lim_{\D\to 0} \D \log q(\D, h) = -\frac{h^2}2 \th(-h) \ , 
\eeq
we obtain
two equations for the jamming line, one for $\D_r$ and the other for $\eta$:
\begin{equation}
\begin{split}
& \Delta_r  = \frac{\widehat{\varphi}_g}{4} \int_{-\infty}^{\eta} dh e^{h} q_{\gamma}(2 \Delta_r; h)  (h-\eta )^2 \ , \\
& 1  = \frac{\widehat{\varphi}_g}{4} \int_{-\infty}^{\eta} dh e^{h} \left[ \frac{\partial}{\partial \Delta_r} q_\gamma \left( 2\Delta_r; h \right)  \right] (h-\eta)^2   \ .
\end{split}
\end{equation}
Note that these equations are equivalent to imposing that
\begin{equation}
H (\Delta_r)= \Delta_r - \frac{\widehat{\varphi}_g}{4} \int_{-\infty}^{\eta} dh e^h (h-\eta)^2 q_\gamma \left( 2 \Delta_r; h \right)
\label{H}
\end{equation}
and its first derivative $H'( \Delta_r)$ both vanish. Their numerical solution is then quite easy.

Note that, according to the replica symmetric (RS) ansatz we use in this paper, in the jamming limit $\wh p \propto \Delta^{-1}$. However, this prediction disagrees with 
the correct scaling solution, $\wh p \simeq \Delta^{-\k}$, regulated by a universal exponent $\k$, 
whose value is exactly predictable only within a full replica symmetry breaking (RSB) approach~\cite{nature}. 
Yet, the RS approximation provides a decent approximation to the location of the jamming critical line. The employment of a full RSB solution might account for the identification of the exact shape of the shear jamming and shear yielding curves in the region where they merge into a critical point, and possibly give rise to a cusp-like trend. This peculiar behavior has been pointed out in recent numerical simulations~\cite{JUZY18} reproducing the stability-reversibility map of hard-sphere glasses.

\subsection{Yielding}
\label{sec:Ydef}

Within mean field theory, a small enough strain applied on a solid gives rise to a static stress, 
up to a point where the applied stress destabilizes the solid: the solution for $(\D,\D_r)$ undergoes a bifurcation and is lost.
We call this point a \emph{yielding transition}~\cite{RUYZ15}, because it is reasonable to conjecture that at larger shear strains the solid yields and starts to flow,
reaching a stationary state in which dynamics is diffusive and the shear stress depends on the shear strain \emph{rate} $\dot\g$. This should be,
however, tested by solving the dynamical mean field equations~\cite{AMZ19}. 

More precisely, the stress $\s(\g)$ increases upon increasing $\g$, overshoots, and then decreases upon approaching the bifurcation (spinodal) point,
where it displays a square root singularity, i.e. $\s(\g) - \s(\g_Y) \propto \sqrt{\gamma_Y-\gamma}$. We determine the yielding transition
line $\g_Y(\h)$, or $\h_Y(\g)$, by following the stress until the singularity is observed.
Note that the appearance of a spinodal corresponds to having a zero \emph{longitudinal} mode in the Hessian matrix of the free energy, 
\beq
M = \begin{pmatrix}
\frac{\partial^2 s_g}{\partial \D \partial \D} & \frac{\partial^2 s_g}{\partial \D \partial \D_r} \\
\frac{\partial^2 s_g}{\partial \D_r \partial \D}&\frac{\partial^2 s_g}{\partial \D_r \partial \D_r}
\end{pmatrix} \ ,
\eeq
which encodes the fluctuations within the replica symmetric sector.
One could determine the yielding point by the condition $\det M=0$, 
but the calculation of $M$ is rather involved and we do not follow this route here. 

Note that comparing this mean-field definition of yielding with experiments or simulations is non trivial, 
and multiple definitions of yielding are used in realistic systems~\cite{BDBDM17}.
Sometimes the yielding point is identified with the maximum of the stress~\cite{Kou12}, 
sometimes with the onset of energy dissipation and the appearance of plastic avalanches~\cite{DHPS16}, 
or with the point where the highest stress drop is observed~\cite{OBBRT18}.
Whether some signature of the mean field spinodal survives in finite dimensions is a currently debated issue~\cite{JPRS16,PRS17,PDW18,OBBRT18,JUZY18}.
Yet, mean field theory provides a qualitatively correct prediction for the shape of the yielding line in the phase diagram~\cite{JUZY18}.
Also in this case, the RS ansatz is sufficient to obtain a correct estimate of the stress overshoot, and of the yielding spinodal point.

\subsection{Gardner transition}
\label{sec:Gdef}

Eqs.~\eqref{iterative-eqs} are iterative equations for the self and the relative mean-square displacements, 
which have been obtained by assuming that the symmetry under permutation of replicas remains unbroken. 
This is not a general assumption and should instead be tested at each state point.
In other words, we should check if the RS solution is actually a stable local minimum of the free energy.

We can then proceed with the computation of the stability matrix. Thanks to the symmetry of the replica indices, the matrix can be simply decomposed in terms of three independent sectors, the \emph{longitudinal} one described above (two scalars), the \emph{anomalous} (vector field) and the \emph{replicon} (tensorial field). The longitudinal mode can generally be associated with the emergence of spinodal points, while the replicon provides information on instabilities that possibly lead to replica symmetry breaking solutions. We refer the interested reader to the specific literature for the computation of these eigenvalues~\cite{DK83,TDP02,RU16}. We only report here the final expression for the replicon mode, 
\begin{equation}\begin{split}
\lambda_\text{R}=  1& -\frac{\widehat{\varphi}_g}{2} \Delta^2 \int_{-\infty}^{\infty} dh e^{h} q_{\gamma}(2 \Delta_r-\Delta; h)   f''(h)^2 \ . 
\end{split}\end{equation}
A positive replicon, $\l_R>0$, defines a stable phase where the RS solution holds. 
Conversely, if the replicon changes sign, the solution becomes unstable and the RS ansatz is no longer correct.
The Gardner transition~\cite{Ga85, KPUZ13, nature} line $\g_{\rm G}(\eta)$, or $\h_{\rm G}(\g)$, 
is then defined by the condition of vanishing replicon eigenvalue, $\l_R=0$.
When $\l_R<0$, each amorphous state, now called {\it basin}, 
is fragmented into a hierarchy of {\it sub-basins}, hence determining a very rough energy landscape~\cite{MPV87, nature, RU16}. 
Describing it requires a more complicated parameterization, via a function $\Delta(x)$ in the interval $x \in \left[0,1\right]$,
called the Parisi order parameter~\cite{MPV87}. 
This function encodes the probability distribution of mean-square displacements between replicas, $\D = \D(X_a,X_b)$,
via the relation $P(\D) = |dx/d\D|$.
The RS solution corresponds to $\D(x)=\D$, $\forall x \in \left[0,1\right]$.

\subsection{Nature of the Gardner transition}
\label{SecGardner}

Note that, while the exact solution for $\D(x)$ above the Gardner transition density requires solving the RSB equations~\cite{RU16},
which goes beyond the scope of this work,
one can evaluate perturbatively its trend close to the Gardner transition, where the replicon is zero~\cite{So85}. We can then compute the \emph{breaking point} $\lambda$,
in whose vicinity $\D(x)$ becomes non-trivial, and the slope 
$\dot{\Delta}(\lambda)$. Using these results, one can identify the universality class of the transition, using only information from the RS calculation.
The breaking point turns out to be:
\begin{equation}
\lambda=  \frac{ \widehat{\varphi}_g \int_{-\infty}^{\infty} dh e^{h} q_{\gamma}(2 \Delta_r-\Delta; h)    f'''(h)^2 } {\frac{4}{\Delta^3} + 2 \widehat{\varphi}_g \int_{-\infty}^{\infty} dh e^h q_{\gamma}(2 \Delta_r-\Delta; h)    f''(h)^3 }  \ .
\end{equation}
while the slope $\dot{\Delta}$, evaluated at the breaking point, is:
\begin{eqnarray}
\dot{\Delta}(\lambda)&=&  \frac{ \frac{4}{\Delta^3} + 2\wh\f_g \int_{-\infty}^{\infty} dh e^h q_{\gamma} (2 \Delta_r-\Delta); h) f''(h)^3 } 
{ \frac{12 \lambda^2}{\Delta^4} -\wh\f_g \int_{-\infty}^{\infty} dh e^h q_{\gamma}(2 \Delta_r-\Delta;h) A(h)} \ ,\nonumber\\
A(h)&=& f''''(h)^2 -12 \lambda f''(h) f'''(h)^2 + 6 \lambda^2 f''(h)^4 \ .
\end{eqnarray}
Based on the values of $\l$ and $\dot\D(\l)$, one can classify the Gardner transition in three universality classes,
keeping in mind that $\D(x)$ must be a decreasing function of $x\in[0,1]$~\cite{nature,RU16},
as follows:
\begin{itemize}
\item 
If the parameter $\lambda$ takes values in $\left[0,1 \right]$, and $\dot \D(\l)<0$,
then one has a {\it bona fide} continuous transition towards a marginally stable fullRSB phase.
\item
If $\lambda\in \left[0,1 \right]$, and $\dot \D(\l)>0$,
then one has a {\it bona fide} continuous transition towards a non-marginal 1RSB phase.
\item
If, instead, $\lambda > 1$, then the transition at $\l_R=0$ is unphysical, and must be preceded by another form of RSB transition, typically a discontinuous RFOT
(exactly as it happens e.g. in the $p$-spin model at low enough external field~\cite{Cavagna1999}
or in the spherical perceptron model~\cite{franz2017}).
\end{itemize}

\begin{figure*}[t]
\includegraphics[width=\textwidth]{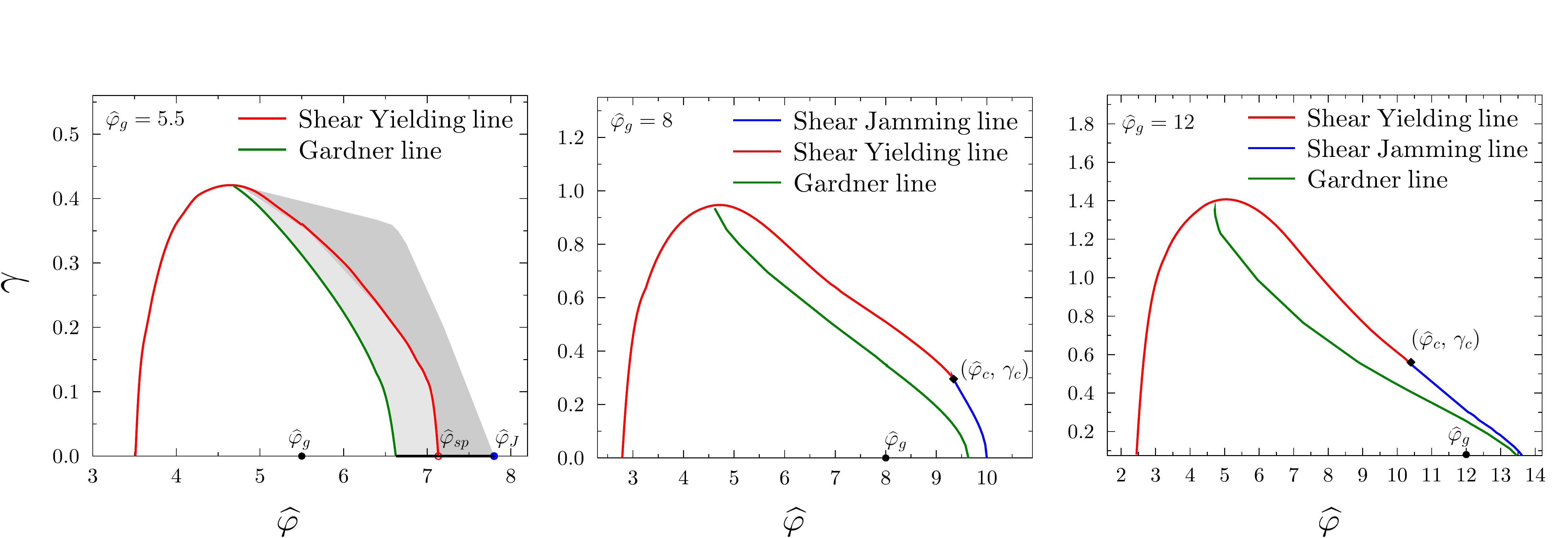}
\caption{Mean-field phase diagrams of a hard-sphere glass 
upon adiabatic following in compression and decompression at density $\wh\f$, and in the presence of a quasi-static shear strain $\gamma$,
for three different preparation densities, $\widehat{\varphi}_g=5.5$, $\widehat{\varphi}_g=8$ and $\widehat{\varphi}_g=12$ respectively. 
 The stability region of the glass is delimited by the shear jamming (blue) and shear yielding (red) lines.
  At sufficiently low $\wh\f_g$, shear jamming disappear and the system can only yield under an applied strain. 
  The Gardner transition line, above which the RS solution is unstable, is shown in green.
  In the left panel, we expect that RSB effects change qualitatively the phase diagram in both gray-shaded regions.}
\label{fig:PD3densities}
\end{figure*}

\section{Results}
\label{SecIV} 

Previous studies~\cite{RUYZ15,UZ17} used the state following method to analyze hard-sphere glasses in the shear jamming and shear yielding regimes. 
Ref.~\cite{UZ17} specifically performed a complete analysis of the phase diagram of a glass prepared at $\widehat{\varphi}_g=8$, under an applied
compression to density $\wh\f \geq \wh\f_g$ and shear strain $\g$.
The aim of the current work is to complete the exploration of the phase diagram at $\wh\f_g=8$, in particular
considering the previously unexplored regime in decompression, and to investigate systematically the dependence of the whole phase diagram on $\wh\f_g$.

\subsection{Glass phase diagram}

In Fig.~\ref{fig:PD3densities} we show the mean-field state following phase diagram in the plane $(\widehat{\varphi},\g)$, 
considering both compression and decompression,
for different values of the initial glass density $\widehat{\varphi}_g=5.5, 8, 12$,
specifically chosen to be representative of the different physical regimes. 
Let us focus first on the phase diagram at $\widehat{\varphi}_g=8$ and later discuss two other instances, respectively at lower and higher densities.

When the glass is compressed at $\g=0$, it exists up to a jamming point at which pressure diverges and $\D\to 0$. When the glass is decompressed
at $\g=0$, it exists up to a melting point where it undergoes a spinodal instability and melts into the liquid. These two transitions thus delimit the region
of existence of the glass at $\g=0$. At $\g>0$, the jamming point extends into a shear jamming line (blue line in Fig.~\ref{fig:PD3densities}),
defined as in section~\ref{sec:Jdef}, while the melting point extends into a shear yielding line (red line in Fig.~\ref{fig:PD3densities}), 
defined as in section~\ref{sec:Ydef}.
The two lines merge into a critical point (black diamond in Fig.~\ref{fig:PD3densities}), whose location $(\widehat{\varphi}_c, \gamma_c)$ 
can be analytically computed considering that there the system is both jammed 
($\Delta=0, \wh p \rightarrow \infty$) and yielded (with a vanishing longitudinal eigenvalue). More precisely, it can be identified by imposing simultaneously that 
Eq.~(\ref{H}) as well as its first and second derivatives vanish, $H(\D_r) = H'\left( \Delta_r\right) = H''\left(\Delta_r \right)=0$.
While it has been shown that the shear jamming and the isotropic (without shear) jamming transition lines fall in the same universality class~\cite{UZ17}, we cannot make any prediction at this level on the critical point, for which two different kinds of instabilities should be taken into account simultaneously.

\begin{figure*}[t]
\includegraphics[width=0.93\textwidth]{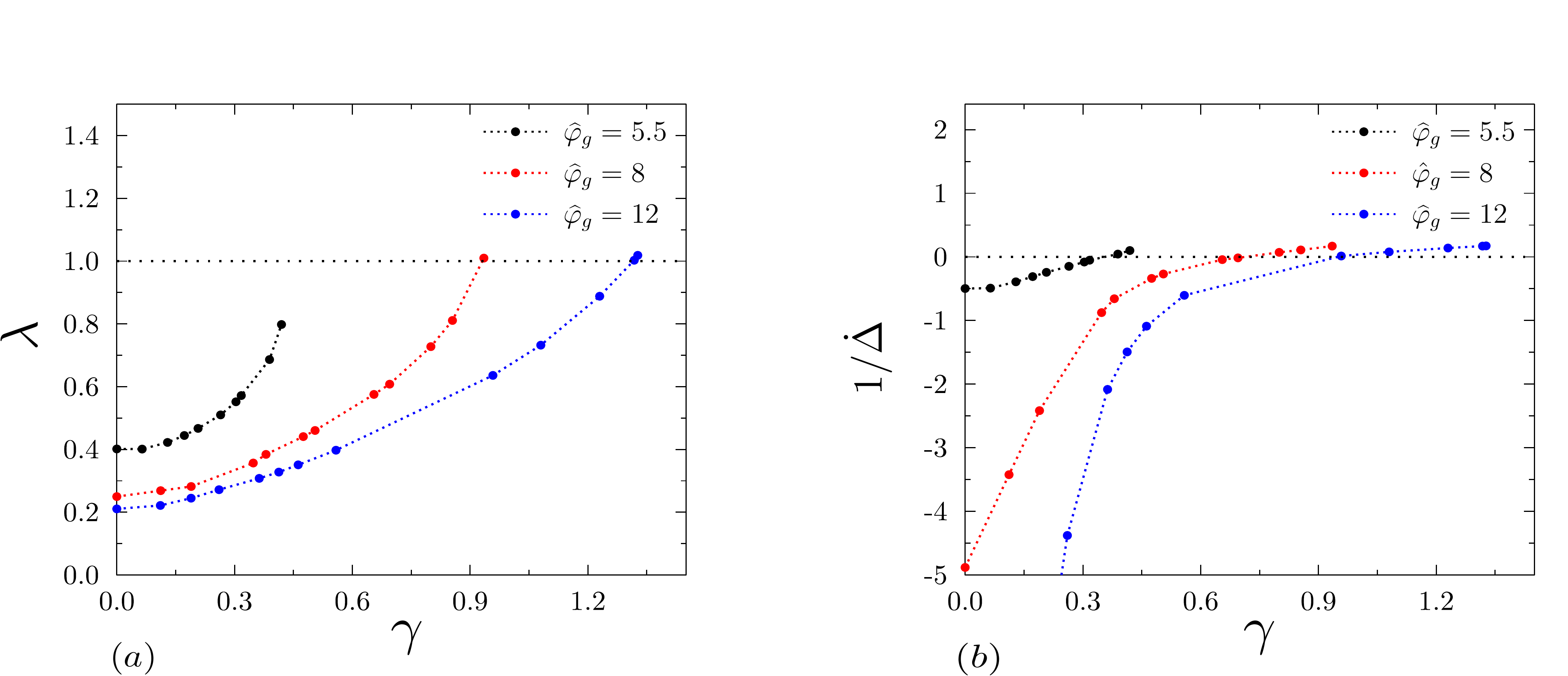}
\caption{(a) Breaking point $\l$ and (b) inverse slope $1/\dot\D(\l)$, as a function of the applied shear strain $\gamma$ along the Gardner transition line,
 for three different increasing values of $\wh\f_g$. }
\label{fig:breakp}
\end{figure*}

Finally, a Gardner transition line is also present (green line in Fig.~\ref{fig:PD3densities}), defined as in Sec.~\ref{sec:Gdef}, separating a stable RS phase (below it) from a RSB phase
(above it). Note that the whole shear jamming line, as well as the critical point, fall into the RSB regime, and their RS value reported here is therefore
only an approximation.
In order to better understand the nature of the Gardner transition line, in Fig.~\ref{fig:breakp} we report the values of the breaking point $\l$ and
the slope $\dot\D(\l)$ computed along the line, as a function of $\g$. At small $\g$, we have $\l<1$ and $\dot\D(\l)<0$, indicating a continuous transition
towards a fullRSB phase, according to the discussion of section~\ref{SecGardner}.
Upon increasing $\g$, the slope $\dot\D(\l)$ diverges and changes sign, while still $\l<1$, indicating that the transition becomes a continuous transition
towards a 1RSB phase. At even larger values of $\g$, $\l$ reaches one, indicating that the transition becomes a RFOT.
This behavior is reminiscent of that of several spin glass models~\cite{Cavagna1999,franz2017}.

We now discuss the two other phase diagrams in Fig.~\ref{fig:PD3densities}, for $\wh\f_g=5.5, 12$.
The phase diagram for $\wh\f_g=12$ mostly differs from the one at $\wh\f_g=8$ for the relative location of the different transitions.
In particular, the shear jamming and the Gardner transition lines appear much closer, with a visible shrinkage of the region delimited by the green and the blue lines as $\gamma \rightarrow 0$. Furthermore, the critical point $(\widehat{\varphi}_c, \gamma_c)$ has now moved at densities smaller than the preparation one, $\wh\f_c<\wh\f_g$, indicating
that an applied shear on the initial equilibrium configuration would lead to shear jamming. The behavior of the breaking point $\l$ and slope $\dot\D(\l)$ are also
qualitatively similar to the $\wh\f_g=8$ case.

The phase diagram for $\wh\f_g=5.5$ presents a more important qualitative difference. In fact, upon decreasing $\wh\f_g$, the critical point moves
to $\g_c=0$ and disappears. The shear jamming line correspondingly disappears, leaving a single shear yielding line that originates from the melting point in
decompression at $\g=0$ and ends into an unphysical spinodal point in compression at $\g=0$~\cite{RUYZ15}. This unphysical spinodal point has been shown
to disappear once RSB is taken into account~\cite{RU16}, and as a consequence the behavior beyond the Gardner transition line (shaded gray area
in Fig.~\ref{fig:PD3densities}) might be qualitatively different once RSB is taken into account. In this case, the value of $\l$ (Fig.~\ref{fig:breakp}) remains always
smaller than one, indicating that the Gardner transition is always continuous.

\begin{figure*}[t]
\includegraphics[width=0.97\textwidth]{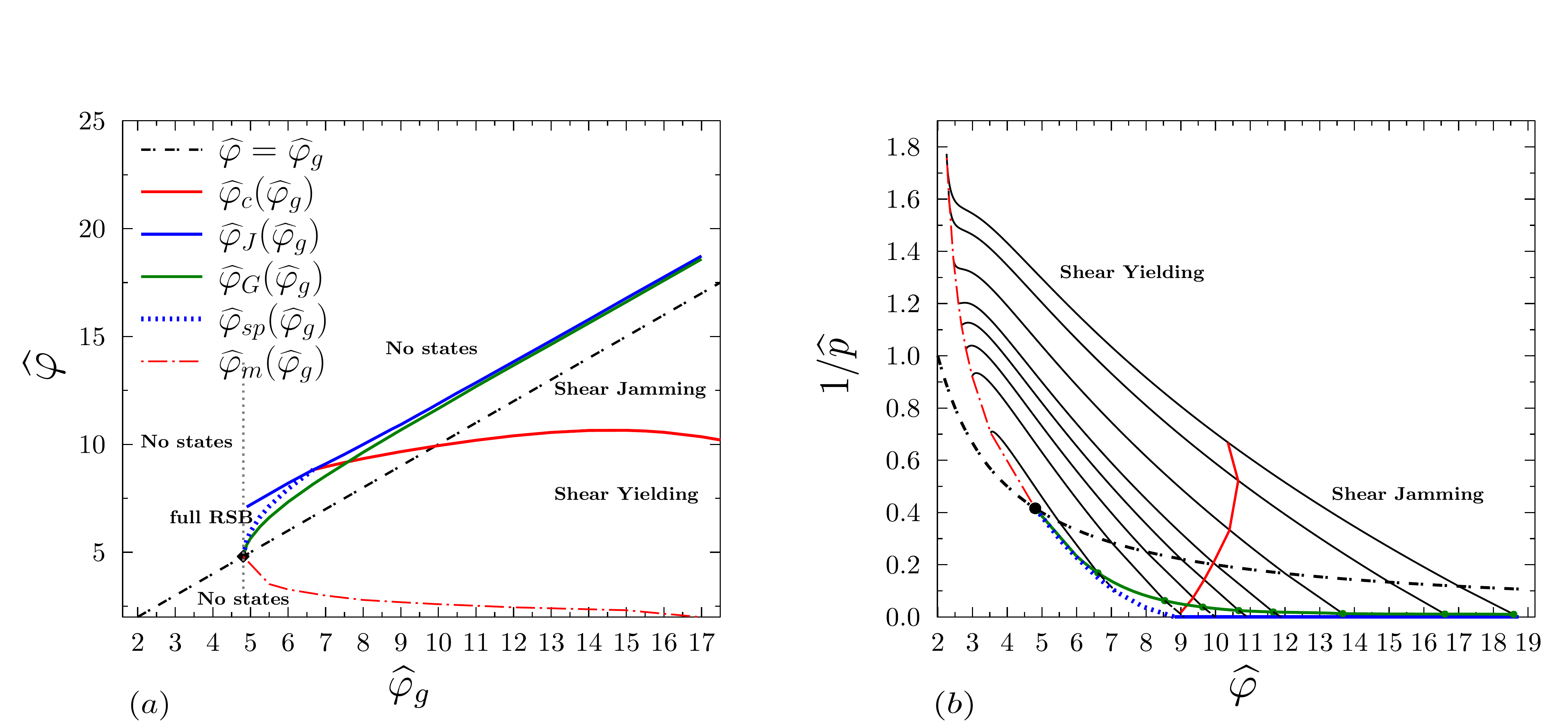}
\caption{
(a) Glass phase diagram in the $(\widehat{\varphi}_g, \widehat {\varphi})$ plane. 
Glasses are prepared in equilibrium on
the bisector line, $\widehat{\varphi}= \widehat{\varphi}_g \geq \wh\f_{\rm d}=4.8067$ (dot-dashed black).
For a given $\wh\f_g$, a glass can be followed in compression up to the jamming transition (blue line/upper curve above which there are no states) or in decompression
down to the melting transition (red dot-dashed line, on the bottom).
Above the Gardner transition (green line, or light gray in the grayscale version), the RS solution becomes unstable towards RSB. In this region, 
for $\widehat{\varphi}_g\leq6.667$, the RS solution predicts an unphysical spinodal (blue dashed line, inside the full RSB region) at which the solution is lost before jamming is reached.
(The continuation of the jamming line for $\widehat{\varphi}_g\leq6.667$ is based on fullRSB data~\cite{RU16}.)
 The critical density $\wh\f_c$ (red line/full gray line in the middle) separates the phase diagram into two regions: above it, the system jams under shear, while below it, it yields. \\
(b) Same phase diagram in the plane of inverse reduced pressure $1/\wh p$ and scaled packing fraction $\widehat{\varphi}$. 
Glasses are prepared on the equilibrium liquid line (dot-dashed black) $\wh p= \widehat{\varphi}_g/2$. 
The dynamical transition density, $\widehat{\varphi}_{\rm d} \simeq 4.8067$, is marked by a full black circle. 
Upon compression or decompression the glasses follow their equation of state (full black lines), reported 
for $\widehat{\varphi}_g=5.5, 7, 8, 9, 10, 12, 15, 17$ (bottom to top), which extends from the melting point (red/light gray dot-dashed line) 
up to jamming (blue line/full line on the horizontal axis) for $\wh\f_g\geq 6.667$ or to the unphysical spinodal (dashed blue/gray line merging with the critical jamming line) for $\wh\f_g\leq 6.667$.
Below the Gardner transition (green/light gray) line, states are unstable towards RSB. The red full line separates the shear yielding and shear jamming regimes.
}
\label{fig:pressure-density}
\end{figure*}

\subsection{Evolution of the glass phase diagram as a function of the preparation density $\widehat{\varphi}_g$}

We now give a broader overview of the evolution of the phase diagram of Fig.~\ref{fig:PD3densities} upon changing $\wh\f_g$.
In Fig.~\ref{fig:pressure-density}(left), 
we plot the critical values of density that define the phase diagram of Fig.~\ref{fig:PD3densities} as a function of $\wh\f_g$.
This $(\widehat{\varphi}_g, \widehat{\varphi})$ phase diagram then shows the evolution with density $\wh\f$ and shear strain $\g$ of a glass prepared at $\wh\f_g$.

For a given $\wh\f_g \geq \wh \f_{\rm d}=4.8067\ldots$, the glass exists in the region of $\wh\f \in [\wh\f_m(\wh\f_g), \wh\f_J(\wh\f_g)]$,
delimited from above by the jamming (blue) line and from below by the melting (red dot-dashed) line.
In between these two lines, we plot a few other important values of density: the Gardner transition point $\wh\f_G$ at $\g=0$ (green line),
the unphysical spinodal point at $\g=0$ (blue dashed line), and the value $\wh\f_c$ (red line) of the density corresponding to the critical point
located at $(\widehat{\varphi}_c, \gamma_c)$.

The interpretation of this phase diagram is the following. The glass state is prepared in equilibrium at $\wh\f = \wh\f_g$ (black dot-dashed line), and
it can be compressed up to $\wh\f_J$ or decompressed down to $\wh\f_m$. Above $\wh\f_G$, the compression brings the system into a RSB phase
and in that case an unphysical spinodal can also be met before jamming. If the system is compressed to $\wh\f <\wh\f_c$ and then strained, shear yielding is
observed, while if it is compressed to $\wh\f>\wh\f_c$, shear jamming is observed.
Note that upon decreasing $\widehat{\varphi}_g$ towards $\approx 6.5$, 
the critical density $\widehat{\varphi}_c$ merges with $\widehat{\varphi}_J$, and the shear jamming regime disappears.

Note that, when the system yields, a stationary flow state with no memory of previous conditions should be reached at long times~\cite{BBK00}.
Fig.~\ref{fig:pressure-density}(left) then suggests that for any density smaller than $\widehat{\varphi}_c(\widehat{\varphi}_g)$, a stationary flow state exists.
It is interesting to observe that the line $\widehat{\varphi}_c(\widehat{\varphi}_g)$ first increases and then decreases, thus showing a single maximum.
The value $\widehat{\varphi}^\text{flow}_J = \max_{\wh\f_g} \widehat{\varphi}_c(\widehat{\varphi}_g) \approx 11$ is then a lower bound to the value
of density for which a hard-sphere system still admits a flowing state.
This result should be confirmed by locating the jamming point under shear strain for hard sphere dynamics, using the dynamical equations derived
in~\cite{AMZ19}.

In Fig.~\ref{fig:pressure-density}(right), we present the same phase diagram in a slightly different representation, in the $(\wh\f, 1/\wh p)$ plane.
The glass is prepared on the equilibrium (dot-dashed black) line, at $\wh\f= \wh\f_g$ and $\wh p = \wh\f_g/2$.
Adiabatic following of the glass state selected in equilibrium by changing its density $\wh\f$ at $\g=0$ 
gives rise to a pressure-density equation of state for each $\wh\f_g$ (black full line). The equation of state is delimited by the 
melting point $(\wh\f_m, 1/\wh p_m)$ at low densities and by the jamming point $(\wh\f_J,0)$ at high densities. 
Note that in the vicinity of the melting point, we have $\wh p - \wh p_m \propto \sqrt{\wh\f - \wh\f_m}$, but the proportionality coefficient changes
sign around $\wh\f_g\sim 10$, from negative at lower $\wh\f_g$ (hence leading to a pressure undershoot before melting) to positive at higher $\wh\f_g$ (hence leading to a monotonic
pressure-density curve).
As in Fig.~\ref{fig:pressure-density}(left),
we report the value of the Gardner point at $\g=0$ (green line), and the value of the critical point, here reported as $(\wh\f_c, 1/\wh p_c)$ where $\wh p_c$ is
the pressure of the glass at $\wh\f_c$ and $\g=0$. 

\section{Conclusions and perspectives}
\label{sec:conc}

In this work we have applied a microscopic mean-field approach to the simplest model of glass former: hard spheres in the $d \rightarrow \infty$ limit.
We obtained the phase diagram of the glass states, followed adiabatically in compression or decompression, and under a quasi-static shear deformation.
Our analysis completes and extends previous analytical studies of rheological properties of hard-sphere systems in high dimensions~\cite{RUYZ15,RU16,UZ17}. 

Remarkably, the phase diagrams shown in Fig.~\ref{fig:PD3densities} are qualitatively similar to those obtained in numerical simulations of three-dimensional systems~\cite{JY17,JUZY18}, for which we also note that for increasing preparation density $\widehat{\varphi}_g$, the values of the yield strain $\gamma_Y$ and the yield stress increase as well. 
The main qualitative difference is the shape of the shear jamming and shear yielding curves in the vicinity of the critical point where they merge, which might be explained by RSB effects.

Indeed,
while our analysis is restricted to the replica symmetric solution, replica symmetry is broken in important portions of the phase diagram, in particular
in the region of high densities close to the shear jamming line and its associated critical point. Within the RS solution, at low preparation density, the shear jamming line disappears and
the shear yielding line extends down to $\g=0$ giving rise to an unphysical spinodal of the glass under compression. 
Clearly, a major direction for future research would be to investigate RSB effects within the state following approach, or even better, to understand what is the dynamical
behavior of glasses followed in the unstable region~\cite{FFR19}.

Another very interesting problem is to identify the exact value of $\widehat{\varphi}^\text{flow}_J$, defined as the density above which no flowing state of hard spheres under
constant shear rate exists. The configurations at $\widehat{\varphi}^\text{flow}_J$ are jammed and likely to be marginal, i.e. characterized by a non-trivial force distribution with universal exponents~\cite{LDW12}. It would be extremely interesting to better understand, within
mean field theory, how the marginal stability condition affects the flow dynamics around the critical density.

\subsection*{Acknowledgments} 

We warmly thank P. Urbani for several insightful interactions on this subject. We also acknowledge Y. Jin and H. Yoshino for useful exchanges and discussions. 
This project has received funding from the European Research Council (ERC) under the European Union’s Horizon 2020 research and innovation program (Grant agreement 723955 - GlassUniversality).

\bibliography{HS}

\begin{thebibliography}{61}%
\makeatletter
\providecommand \@ifxundefined [1]{%
 \@ifx{#1\undefined}
}%
\providecommand \@ifnum [1]{%
 \ifnum #1\expandafter \@firstoftwo
 \else \expandafter \@secondoftwo
 \fi
}%
\providecommand \@ifx [1]{%
 \ifx #1\expandafter \@firstoftwo
 \else \expandafter \@secondoftwo
 \fi
}%
\providecommand \natexlab [1]{#1}%
\providecommand \enquote  [1]{``#1''}%
\providecommand \bibnamefont  [1]{#1}%
\providecommand \bibfnamefont [1]{#1}%
\providecommand \citenamefont [1]{#1}%
\providecommand \href@noop [0]{\@secondoftwo}%
\providecommand \href [0]{\begingroup \@sanitize@url \@href}%
\providecommand \@href[1]{\@@startlink{#1}\@@href}%
\providecommand \@@href[1]{\endgroup#1\@@endlink}%
\providecommand \@sanitize@url [0]{\catcode `\\12\catcode `\$12\catcode
  `\&12\catcode `\#12\catcode `\^12\catcode `\_12\catcode `\%12\relax}%
\providecommand \@@startlink[1]{}%
\providecommand \@@endlink[0]{}%
\providecommand \url  [0]{\begingroup\@sanitize@url \@url }%
\providecommand \@url [1]{\endgroup\@href {#1}{\urlprefix }}%
\providecommand \urlprefix  [0]{URL }%
\providecommand \Eprint [0]{\href }%
\providecommand \doibase [0]{http://dx.doi.org/}%
\providecommand \selectlanguage [0]{\@gobble}%
\providecommand \bibinfo  [0]{\@secondoftwo}%
\providecommand \bibfield  [0]{\@secondoftwo}%
\providecommand \translation [1]{[#1]}%
\providecommand \BibitemOpen [0]{}%
\providecommand \bibitemStop [0]{}%
\providecommand \bibitemNoStop [0]{.\EOS\space}%
\providecommand \EOS [0]{\spacefactor3000\relax}%
\providecommand \BibitemShut  [1]{\csname bibitem#1\endcsname}%
\let\auto@bib@innerbib\@empty
\bibitem [{\citenamefont {Rodney}\ \emph {et~al.}(2011)\citenamefont {Rodney},
  \citenamefont {Tanguy},\ and\ \citenamefont {Vandembroucq}}]{RTV11}%
  \BibitemOpen
  \bibfield  {author} {\bibinfo {author} {\bibfnamefont {D.}~\bibnamefont
  {Rodney}}, \bibinfo {author} {\bibfnamefont {A.}~\bibnamefont {Tanguy}}, \
  and\ \bibinfo {author} {\bibfnamefont {D.}~\bibnamefont {Vandembroucq}},\
  }\href@noop {} {\bibfield  {journal} {\bibinfo  {journal} {Modelling and
  Simulation in Materials Science and Engineering}\ }\textbf {\bibinfo {volume}
  {19}},\ \bibinfo {pages} {083001} (\bibinfo {year} {2011})}\BibitemShut
  {NoStop}%
\bibitem [{\citenamefont {Bonn}\ \emph {et~al.}(2017)\citenamefont {Bonn},
  \citenamefont {Denn}, \citenamefont {Berthier}, \citenamefont {Divoux},\ and\
  \citenamefont {Manneville}}]{BDBDM17}%
  \BibitemOpen
  \bibfield  {author} {\bibinfo {author} {\bibfnamefont {D.}~\bibnamefont
  {Bonn}}, \bibinfo {author} {\bibfnamefont {M.~M.}\ \bibnamefont {Denn}},
  \bibinfo {author} {\bibfnamefont {L.}~\bibnamefont {Berthier}}, \bibinfo
  {author} {\bibfnamefont {T.}~\bibnamefont {Divoux}}, \ and\ \bibinfo {author}
  {\bibfnamefont {S.}~\bibnamefont {Manneville}},\ }\href@noop {} {\bibfield
  {journal} {\bibinfo  {journal} {Reviews of Modern Physics}\ }\textbf
  {\bibinfo {volume} {89}},\ \bibinfo {pages} {035005} (\bibinfo {year}
  {2017})}\BibitemShut {NoStop}%
\bibitem [{\citenamefont {Ozawa}\ \emph {et~al.}(2018)\citenamefont {Ozawa},
  \citenamefont {Berthier}, \citenamefont {Biroli}, \citenamefont {Rosso},\
  and\ \citenamefont {Tarjus}}]{OBBRT18}%
  \BibitemOpen
  \bibfield  {author} {\bibinfo {author} {\bibfnamefont {M.}~\bibnamefont
  {Ozawa}}, \bibinfo {author} {\bibfnamefont {L.}~\bibnamefont {Berthier}},
  \bibinfo {author} {\bibfnamefont {G.}~\bibnamefont {Biroli}}, \bibinfo
  {author} {\bibfnamefont {A.}~\bibnamefont {Rosso}}, \ and\ \bibinfo {author}
  {\bibfnamefont {G.}~\bibnamefont {Tarjus}},\ }\href@noop {} {\bibfield
  {journal} {\bibinfo  {journal} {Proceedings of the National Academy of
  Sciences}\ }\textbf {\bibinfo {volume} {115}},\ \bibinfo {pages} {6656}
  (\bibinfo {year} {2018})}\BibitemShut {NoStop}%
\bibitem [{\citenamefont {Liu}\ and\ \citenamefont {Nagel}(2010)}]{LN10}%
  \BibitemOpen
  \bibfield  {author} {\bibinfo {author} {\bibfnamefont {A.~J.}\ \bibnamefont
  {Liu}}\ and\ \bibinfo {author} {\bibfnamefont {S.~R.}\ \bibnamefont
  {Nagel}},\ }\href@noop {} {\bibfield  {journal} {\bibinfo  {journal} {Annual
  Reviews of Condensed Matter Physics}\ }\textbf {\bibinfo {volume} {1}},\
  \bibinfo {pages} {347} (\bibinfo {year} {2010})}\BibitemShut {NoStop}%
\bibitem [{\citenamefont {M{\"u}ller}\ and\ \citenamefont
  {Wyart}(2015)}]{MW15}%
  \BibitemOpen
  \bibfield  {author} {\bibinfo {author} {\bibfnamefont {M.}~\bibnamefont
  {M{\"u}ller}}\ and\ \bibinfo {author} {\bibfnamefont {M.}~\bibnamefont
  {Wyart}},\ }\href@noop {} {\bibfield  {journal} {\bibinfo  {journal} {Annual
  Reviews of Condensed Matter Physics}\ }\textbf {\bibinfo {volume} {6}},\
  \bibinfo {pages} {177} (\bibinfo {year} {2015})}\BibitemShut {NoStop}%
\bibitem [{\citenamefont {Wyart}\ \emph {et~al.}(2005)\citenamefont {Wyart},
  \citenamefont {Nagel},\ and\ \citenamefont {Witten}}]{WNW05}%
  \BibitemOpen
  \bibfield  {author} {\bibinfo {author} {\bibfnamefont {M.}~\bibnamefont
  {Wyart}}, \bibinfo {author} {\bibfnamefont {S.}~\bibnamefont {Nagel}}, \ and\
  \bibinfo {author} {\bibfnamefont {T.}~\bibnamefont {Witten}},\ }\href@noop {}
  {\bibfield  {journal} {\bibinfo  {journal} {Europhysics Letters}\ }\textbf
  {\bibinfo {volume} {72}},\ \bibinfo {pages} {486} (\bibinfo {year}
  {2005})}\BibitemShut {NoStop}%
\bibitem [{\citenamefont {Berthier}\ \emph {et~al.}(2016)\citenamefont
  {Berthier}, \citenamefont {Charbonneau}, \citenamefont {Jin}, \citenamefont
  {Parisi}, \citenamefont {Seoane},\ and\ \citenamefont {Zamponi}}]{BCJPSZ16}%
  \BibitemOpen
  \bibfield  {author} {\bibinfo {author} {\bibfnamefont {L.}~\bibnamefont
  {Berthier}}, \bibinfo {author} {\bibfnamefont {P.}~\bibnamefont
  {Charbonneau}}, \bibinfo {author} {\bibfnamefont {Y.}~\bibnamefont {Jin}},
  \bibinfo {author} {\bibfnamefont {G.}~\bibnamefont {Parisi}}, \bibinfo
  {author} {\bibfnamefont {B.}~\bibnamefont {Seoane}}, \ and\ \bibinfo {author}
  {\bibfnamefont {F.}~\bibnamefont {Zamponi}},\ }\href {\doibase
  10.1073/pnas.1607730113} {\bibfield  {journal} {\bibinfo  {journal}
  {Proceedings of the National Academy of Sciences}\ }\textbf {\bibinfo
  {volume} {113}},\ \bibinfo {pages} {8397} (\bibinfo {year}
  {2016})}\BibitemShut {NoStop}%
\bibitem [{\citenamefont {Kirkpatrick}\ and\ \citenamefont
  {Wolynes}(1987)}]{KW87}%
  \BibitemOpen
  \bibfield  {author} {\bibinfo {author} {\bibfnamefont {T.~R.}\ \bibnamefont
  {Kirkpatrick}}\ and\ \bibinfo {author} {\bibfnamefont {P.~G.}\ \bibnamefont
  {Wolynes}},\ }\href {\doibase 10.1103/PhysRevA.35.3072} {\bibfield  {journal}
  {\bibinfo  {journal} {Phys. Rev. A}\ }\textbf {\bibinfo {volume} {35}},\
  \bibinfo {pages} {3072} (\bibinfo {year} {1987})}\BibitemShut {NoStop}%
\bibitem [{\citenamefont {Kirkpatrick}\ and\ \citenamefont
  {Thirumalai}(1987)}]{KT87}%
  \BibitemOpen
  \bibfield  {author} {\bibinfo {author} {\bibfnamefont {T.~R.}\ \bibnamefont
  {Kirkpatrick}}\ and\ \bibinfo {author} {\bibfnamefont {D.}~\bibnamefont
  {Thirumalai}},\ }\href@noop {} {\bibfield  {journal} {\bibinfo  {journal}
  {Phys. Rev. Lett.}\ }\textbf {\bibinfo {volume} {58}},\ \bibinfo {pages}
  {2091} (\bibinfo {year} {1987})}\BibitemShut {NoStop}%
\bibitem [{\citenamefont {Kirkpatrick}\ \emph {et~al.}(1989)\citenamefont
  {Kirkpatrick}, \citenamefont {Thirumalai},\ and\ \citenamefont
  {Wolynes}}]{KTW89}%
  \BibitemOpen
  \bibfield  {author} {\bibinfo {author} {\bibfnamefont {T.~R.}\ \bibnamefont
  {Kirkpatrick}}, \bibinfo {author} {\bibfnamefont {D.}~\bibnamefont
  {Thirumalai}}, \ and\ \bibinfo {author} {\bibfnamefont {P.~G.}\ \bibnamefont
  {Wolynes}},\ }\href {\doibase 10.1103/PhysRevA.40.1045} {\bibfield  {journal}
  {\bibinfo  {journal} {Phys. Rev. A}\ }\textbf {\bibinfo {volume} {40}},\
  \bibinfo {pages} {1045} (\bibinfo {year} {1989})}\BibitemShut {NoStop}%
\bibitem [{\citenamefont {Singh}\ \emph {et~al.}(1985)\citenamefont {Singh},
  \citenamefont {Stoessel},\ and\ \citenamefont {Wolynes}}]{SSW85}%
  \BibitemOpen
  \bibfield  {author} {\bibinfo {author} {\bibfnamefont {Y.}~\bibnamefont
  {Singh}}, \bibinfo {author} {\bibfnamefont {J.~P.}\ \bibnamefont {Stoessel}},
  \ and\ \bibinfo {author} {\bibfnamefont {P.~G.}\ \bibnamefont {Wolynes}},\
  }\href {\doibase 10.1103/PhysRevLett.54.1059} {\bibfield  {journal} {\bibinfo
   {journal} {Phys. Rev. Lett.}\ }\textbf {\bibinfo {volume} {54}},\ \bibinfo
  {pages} {1059} (\bibinfo {year} {1985})}\BibitemShut {NoStop}%
\bibitem [{\citenamefont {Rainone}\ \emph {et~al.}(2015)\citenamefont
  {Rainone}, \citenamefont {Urbani}, \citenamefont {Yoshino},\ and\
  \citenamefont {Zamponi}}]{RUYZ15}%
  \BibitemOpen
  \bibfield  {author} {\bibinfo {author} {\bibfnamefont {C.}~\bibnamefont
  {Rainone}}, \bibinfo {author} {\bibfnamefont {P.}~\bibnamefont {Urbani}},
  \bibinfo {author} {\bibfnamefont {H.}~\bibnamefont {Yoshino}}, \ and\
  \bibinfo {author} {\bibfnamefont {F.}~\bibnamefont {Zamponi}},\ }\href@noop
  {} {\bibfield  {journal} {\bibinfo  {journal} {Physical Review Letters}\
  }\textbf {\bibinfo {volume} {114}},\ \bibinfo {pages} {015701} (\bibinfo
  {year} {2015})}\BibitemShut {NoStop}%
\bibitem [{\citenamefont {Franz}\ and\ \citenamefont {Parisi}(1995)}]{FP95}%
  \BibitemOpen
  \bibfield  {author} {\bibinfo {author} {\bibfnamefont {S.}~\bibnamefont
  {Franz}}\ and\ \bibinfo {author} {\bibfnamefont {G.}~\bibnamefont {Parisi}},\
  }\href@noop {} {\bibfield  {journal} {\bibinfo  {journal} {Journal de
  Physique I}\ }\textbf {\bibinfo {volume} {5}},\ \bibinfo {pages} {1401}
  (\bibinfo {year} {1995})}\BibitemShut {NoStop}%
\bibitem [{\citenamefont {Franz}\ and\ \citenamefont {Parisi}(1997)}]{FP97}%
  \BibitemOpen
  \bibfield  {author} {\bibinfo {author} {\bibfnamefont {S.}~\bibnamefont
  {Franz}}\ and\ \bibinfo {author} {\bibfnamefont {G.}~\bibnamefont {Parisi}},\
  }\href {\doibase 10.1103/PhysRevLett.79.2486} {\bibfield  {journal} {\bibinfo
   {journal} {Phys. Rev. Lett.}\ }\textbf {\bibinfo {volume} {79}},\ \bibinfo
  {pages} {2486} (\bibinfo {year} {1997})}\BibitemShut {NoStop}%
\bibitem [{\citenamefont {Barrat}\ \emph {et~al.}(1997)\citenamefont {Barrat},
  \citenamefont {Franz},\ and\ \citenamefont {Parisi}}]{BFP97}%
  \BibitemOpen
  \bibfield  {author} {\bibinfo {author} {\bibfnamefont {A.}~\bibnamefont
  {Barrat}}, \bibinfo {author} {\bibfnamefont {S.}~\bibnamefont {Franz}}, \
  and\ \bibinfo {author} {\bibfnamefont {G.}~\bibnamefont {Parisi}},\ }\href
  {http://stacks.iop.org/0305-4470/30/5593} {\bibfield  {journal} {\bibinfo
  {journal} {Journal of Physics A: Mathematical and General}\ }\textbf
  {\bibinfo {volume} {30}},\ \bibinfo {pages} {5593} (\bibinfo {year}
  {1997})}\BibitemShut {NoStop}%
\bibitem [{\citenamefont {Krzakala}\ and\ \citenamefont
  {Zdeborov{\'a}}(2010)}]{KZ10}%
  \BibitemOpen
  \bibfield  {author} {\bibinfo {author} {\bibfnamefont {F.}~\bibnamefont
  {Krzakala}}\ and\ \bibinfo {author} {\bibfnamefont {L.}~\bibnamefont
  {Zdeborov{\'a}}},\ }\href@noop {} {\bibfield  {journal} {\bibinfo  {journal}
  {EPL}\ }\textbf {\bibinfo {volume} {90}},\ \bibinfo {pages} {66002} (\bibinfo
  {year} {2010})}\BibitemShut {NoStop}%
\bibitem [{\citenamefont {Zdeborov\'a}\ and\ \citenamefont
  {Krzakala}(2010)}]{KZ10b}%
  \BibitemOpen
  \bibfield  {author} {\bibinfo {author} {\bibfnamefont {L.}~\bibnamefont
  {Zdeborov\'a}}\ and\ \bibinfo {author} {\bibfnamefont {F.}~\bibnamefont
  {Krzakala}},\ }\href@noop {} {\bibfield  {journal} {\bibinfo  {journal}
  {Phys.~Rev.~B}\ }\textbf {\bibinfo {volume} {81}},\ \bibinfo {pages} {224205}
  (\bibinfo {year} {2010})}\BibitemShut {NoStop}%
\bibitem [{\citenamefont {Krzakala}\ and\ \citenamefont
  {Zdeborov{\'a}}(2013)}]{KZ13}%
  \BibitemOpen
  \bibfield  {author} {\bibinfo {author} {\bibfnamefont {F.}~\bibnamefont
  {Krzakala}}\ and\ \bibinfo {author} {\bibfnamefont {L.}~\bibnamefont
  {Zdeborov{\'a}}},\ }\href@noop {} {\bibfield  {journal} {\bibinfo  {journal}
  {Journal of Physics: Conference Series}\ }\textbf {\bibinfo {volume} {473}},\
  \bibinfo {pages} {12022} (\bibinfo {year} {2013})}\BibitemShut {NoStop}%
\bibitem [{\citenamefont {Charbonneau}\ \emph {et~al.}(2017)\citenamefont
  {Charbonneau}, \citenamefont {Kurchan}, \citenamefont {Parisi}, \citenamefont
  {Urbani},\ and\ \citenamefont {Zamponi}}]{CKPUZ17}%
  \BibitemOpen
  \bibfield  {author} {\bibinfo {author} {\bibfnamefont {P.}~\bibnamefont
  {Charbonneau}}, \bibinfo {author} {\bibfnamefont {J.}~\bibnamefont
  {Kurchan}}, \bibinfo {author} {\bibfnamefont {G.}~\bibnamefont {Parisi}},
  \bibinfo {author} {\bibfnamefont {P.}~\bibnamefont {Urbani}}, \ and\ \bibinfo
  {author} {\bibfnamefont {F.}~\bibnamefont {Zamponi}},\ }\href@noop {}
  {\bibfield  {journal} {\bibinfo  {journal} {Annual Review of Condensed Matter
  Physics}\ }\textbf {\bibinfo {volume} {8}},\ \bibinfo {pages} {265} (\bibinfo
  {year} {2017})}\BibitemShut {NoStop}%
\bibitem [{\citenamefont {M{\'e}zard}\ and\ \citenamefont
  {Parisi}(1996)}]{MP96}%
  \BibitemOpen
  \bibfield  {author} {\bibinfo {author} {\bibfnamefont {M.}~\bibnamefont
  {M{\'e}zard}}\ and\ \bibinfo {author} {\bibfnamefont {G.}~\bibnamefont
  {Parisi}},\ }\href@noop {} {\bibfield  {journal} {\bibinfo  {journal}
  {Journal of Physics A: Mathematical and General}\ }\textbf {\bibinfo {volume}
  {29}},\ \bibinfo {pages} {6515} (\bibinfo {year} {1996})}\BibitemShut
  {NoStop}%
\bibitem [{\citenamefont {Cardenas}\ \emph {et~al.}(1998)\citenamefont
  {Cardenas}, \citenamefont {Franz},\ and\ \citenamefont {Parisi}}]{CFP98}%
  \BibitemOpen
  \bibfield  {author} {\bibinfo {author} {\bibfnamefont {M.}~\bibnamefont
  {Cardenas}}, \bibinfo {author} {\bibfnamefont {S.}~\bibnamefont {Franz}}, \
  and\ \bibinfo {author} {\bibfnamefont {G.}~\bibnamefont {Parisi}},\ }\href
  {http://stacks.iop.org/0305-4470/31/L163} {\bibfield  {journal} {\bibinfo
  {journal} {Journal of Physics A: Mathematical and General}\ }\textbf
  {\bibinfo {volume} {31}},\ \bibinfo {pages} {L163} (\bibinfo {year}
  {1998})}\BibitemShut {NoStop}%
\bibitem [{\citenamefont {Yoshino}\ and\ \citenamefont
  {M\'ezard}(2010)}]{YM10}%
  \BibitemOpen
  \bibfield  {author} {\bibinfo {author} {\bibfnamefont {H.}~\bibnamefont
  {Yoshino}}\ and\ \bibinfo {author} {\bibfnamefont {M.}~\bibnamefont
  {M\'ezard}},\ }\href@noop {} {\bibfield  {journal} {\bibinfo  {journal}
  {Phys. Rev. Lett.}\ }\textbf {\bibinfo {volume} {105}},\ \bibinfo {pages}
  {015504} (\bibinfo {year} {2010})}\BibitemShut {NoStop}%
\bibitem [{\citenamefont {Parisi}\ and\ \citenamefont {Zamponi}(2010)}]{PZ10}%
  \BibitemOpen
  \bibfield  {author} {\bibinfo {author} {\bibfnamefont {G.}~\bibnamefont
  {Parisi}}\ and\ \bibinfo {author} {\bibfnamefont {F.}~\bibnamefont
  {Zamponi}},\ }\href@noop {} {\bibfield  {journal} {\bibinfo  {journal} {Rev.
  Mod. Phys.}\ }\textbf {\bibinfo {volume} {82}},\ \bibinfo {pages} {789}
  (\bibinfo {year} {2010})}\BibitemShut {NoStop}%
\bibitem [{\citenamefont {Urbani}\ and\ \citenamefont {Zamponi}(2017)}]{UZ17}%
  \BibitemOpen
  \bibfield  {author} {\bibinfo {author} {\bibfnamefont {P.}~\bibnamefont
  {Urbani}}\ and\ \bibinfo {author} {\bibfnamefont {F.}~\bibnamefont
  {Zamponi}},\ }\href@noop {} {\bibfield  {journal} {\bibinfo  {journal}
  {Physical review letters}\ }\textbf {\bibinfo {volume} {118}},\ \bibinfo
  {pages} {038001} (\bibinfo {year} {2017})}\BibitemShut {NoStop}%
\bibitem [{\citenamefont {Yoshino}\ and\ \citenamefont {Zamponi}(2014)}]{YZ14}%
  \BibitemOpen
  \bibfield  {author} {\bibinfo {author} {\bibfnamefont {H.}~\bibnamefont
  {Yoshino}}\ and\ \bibinfo {author} {\bibfnamefont {F.}~\bibnamefont
  {Zamponi}},\ }\href@noop {} {\bibfield  {journal} {\bibinfo  {journal} {Phys.
  Rev. E}\ }\textbf {\bibinfo {volume} {90}},\ \bibinfo {pages} {022302}
  (\bibinfo {year} {2014})}\BibitemShut {NoStop}%
\bibitem [{\citenamefont {Gardner}(1985)}]{Ga85}%
  \BibitemOpen
  \bibfield  {author} {\bibinfo {author} {\bibfnamefont {E.}~\bibnamefont
  {Gardner}},\ }\href@noop {} {\bibfield  {journal} {\bibinfo  {journal}
  {Nuclear Physics B}\ }\textbf {\bibinfo {volume} {257}},\ \bibinfo {pages}
  {747} (\bibinfo {year} {1985})}\BibitemShut {NoStop}%
\bibitem [{\citenamefont {Kurchan}\ \emph {et~al.}(2013)\citenamefont
  {Kurchan}, \citenamefont {Parisi}, \citenamefont {Urbani},\ and\
  \citenamefont {Zamponi}}]{KPUZ13}%
  \BibitemOpen
  \bibfield  {author} {\bibinfo {author} {\bibfnamefont {J.}~\bibnamefont
  {Kurchan}}, \bibinfo {author} {\bibfnamefont {G.}~\bibnamefont {Parisi}},
  \bibinfo {author} {\bibfnamefont {P.}~\bibnamefont {Urbani}}, \ and\ \bibinfo
  {author} {\bibfnamefont {F.}~\bibnamefont {Zamponi}},\ }\href@noop {}
  {\bibfield  {journal} {\bibinfo  {journal} {J. Phys. Chem. B}\ }\textbf
  {\bibinfo {volume} {117}},\ \bibinfo {pages} {12979} (\bibinfo {year}
  {2013})}\BibitemShut {NoStop}%
\bibitem [{\citenamefont {Charbonneau}\ \emph {et~al.}(2014)\citenamefont
  {Charbonneau}, \citenamefont {Kurchan}, \citenamefont {Parisi}, \citenamefont
  {Urbani},\ and\ \citenamefont {Zamponi}}]{nature}%
  \BibitemOpen
  \bibfield  {author} {\bibinfo {author} {\bibfnamefont {P.}~\bibnamefont
  {Charbonneau}}, \bibinfo {author} {\bibfnamefont {J.}~\bibnamefont
  {Kurchan}}, \bibinfo {author} {\bibfnamefont {G.}~\bibnamefont {Parisi}},
  \bibinfo {author} {\bibfnamefont {P.}~\bibnamefont {Urbani}}, \ and\ \bibinfo
  {author} {\bibfnamefont {F.}~\bibnamefont {Zamponi}},\ }\href@noop {}
  {\bibfield  {journal} {\bibinfo  {journal} {Nature Communications}\ }\textbf
  {\bibinfo {volume} {5}},\ \bibinfo {pages} {3725} (\bibinfo {year}
  {2014})}\BibitemShut {NoStop}%
\bibitem [{\citenamefont {Rainone}\ and\ \citenamefont {Urbani}(2016)}]{RU16}%
  \BibitemOpen
  \bibfield  {author} {\bibinfo {author} {\bibfnamefont {C.}~\bibnamefont
  {Rainone}}\ and\ \bibinfo {author} {\bibfnamefont {P.}~\bibnamefont
  {Urbani}},\ }\href@noop {} {\bibfield  {journal} {\bibinfo  {journal}
  {Journal of Statistical Mechanics: Theory and Experiment}\ }\textbf {\bibinfo
  {volume} {2016}},\ \bibinfo {pages} {053302} (\bibinfo {year}
  {2016})}\BibitemShut {NoStop}%
\bibitem [{\citenamefont {Biroli}\ and\ \citenamefont {Urbani}(2016)}]{BU16}%
  \BibitemOpen
  \bibfield  {author} {\bibinfo {author} {\bibfnamefont {G.}~\bibnamefont
  {Biroli}}\ and\ \bibinfo {author} {\bibfnamefont {P.}~\bibnamefont
  {Urbani}},\ }\href@noop {} {\bibfield  {journal} {\bibinfo  {journal} {Nature
  Physics}\ } (\bibinfo {year} {2016})}\BibitemShut {NoStop}%
\bibitem [{\citenamefont {Biroli}\ and\ \citenamefont {Urbani}(2018)}]{BU18}%
  \BibitemOpen
  \bibfield  {author} {\bibinfo {author} {\bibfnamefont {G.}~\bibnamefont
  {Biroli}}\ and\ \bibinfo {author} {\bibfnamefont {P.}~\bibnamefont
  {Urbani}},\ }\href@noop {} {\bibfield  {journal} {\bibinfo  {journal}
  {SciPost Physics}\ }\textbf {\bibinfo {volume} {4}},\ \bibinfo {pages} {020}
  (\bibinfo {year} {2018})}\BibitemShut {NoStop}%
\bibitem [{\citenamefont {Franz}\ and\ \citenamefont {Spigler}(2017)}]{FS17}%
  \BibitemOpen
  \bibfield  {author} {\bibinfo {author} {\bibfnamefont {S.}~\bibnamefont
  {Franz}}\ and\ \bibinfo {author} {\bibfnamefont {S.}~\bibnamefont
  {Spigler}},\ }\href@noop {} {\bibfield  {journal} {\bibinfo  {journal}
  {Physical Review E}\ }\textbf {\bibinfo {volume} {95}},\ \bibinfo {pages}
  {022139} (\bibinfo {year} {2017})}\BibitemShut {NoStop}%
\bibitem [{\citenamefont {Sellitto}\ and\ \citenamefont
  {Zamponi}(2013)}]{SZ13}%
  \BibitemOpen
  \bibfield  {author} {\bibinfo {author} {\bibfnamefont {M.}~\bibnamefont
  {Sellitto}}\ and\ \bibinfo {author} {\bibfnamefont {F.}~\bibnamefont
  {Zamponi}},\ }\href@noop {} {\bibfield  {journal} {\bibinfo  {journal} {EPL
  (Europhysics Letters)}\ }\textbf {\bibinfo {volume} {103}},\ \bibinfo {pages}
  {46005} (\bibinfo {year} {2013})}\BibitemShut {NoStop}%
\bibitem [{\citenamefont {Altieri}\ \emph {et~al.}(2018)\citenamefont
  {Altieri}, \citenamefont {Urbani},\ and\ \citenamefont {Zamponi}}]{AUZ18}%
  \BibitemOpen
  \bibfield  {author} {\bibinfo {author} {\bibfnamefont {A.}~\bibnamefont
  {Altieri}}, \bibinfo {author} {\bibfnamefont {P.}~\bibnamefont {Urbani}}, \
  and\ \bibinfo {author} {\bibfnamefont {F.}~\bibnamefont {Zamponi}},\
  }\href@noop {} {\bibfield  {journal} {\bibinfo  {journal} {Phys. Rev. Lett.}\
  }\textbf {\bibinfo {volume} {121}},\ \bibinfo {pages} {185503} (\bibinfo
  {year} {2018})}\BibitemShut {NoStop}%
\bibitem [{\citenamefont {Charbonneau}\ \emph {et~al.}(2012)\citenamefont
  {Charbonneau}, \citenamefont {Corwin}, \citenamefont {Parisi},\ and\
  \citenamefont {Zamponi}}]{CCPZ12}%
  \BibitemOpen
  \bibfield  {author} {\bibinfo {author} {\bibfnamefont {P.}~\bibnamefont
  {Charbonneau}}, \bibinfo {author} {\bibfnamefont {E.~I.}\ \bibnamefont
  {Corwin}}, \bibinfo {author} {\bibfnamefont {G.}~\bibnamefont {Parisi}}, \
  and\ \bibinfo {author} {\bibfnamefont {F.}~\bibnamefont {Zamponi}},\ }\href
  {\doibase 10.1103/PhysRevLett.109.205501} {\bibfield  {journal} {\bibinfo
  {journal} {Phys. Rev. Lett.}\ }\textbf {\bibinfo {volume} {109}},\ \bibinfo
  {pages} {205501} (\bibinfo {year} {2012})}\BibitemShut {NoStop}%
\bibitem [{\citenamefont {Charbonneau}\ \emph {et~al.}(2015)\citenamefont
  {Charbonneau}, \citenamefont {Corwin}, \citenamefont {Parisi},\ and\
  \citenamefont {Zamponi}}]{CCPZ15}%
  \BibitemOpen
  \bibfield  {author} {\bibinfo {author} {\bibfnamefont {P.}~\bibnamefont
  {Charbonneau}}, \bibinfo {author} {\bibfnamefont {E.~I.}\ \bibnamefont
  {Corwin}}, \bibinfo {author} {\bibfnamefont {G.}~\bibnamefont {Parisi}}, \
  and\ \bibinfo {author} {\bibfnamefont {F.}~\bibnamefont {Zamponi}},\
  }\href@noop {} {\bibfield  {journal} {\bibinfo  {journal} {Physical review
  letters}\ }\textbf {\bibinfo {volume} {114}},\ \bibinfo {pages} {125504}
  (\bibinfo {year} {2015})}\BibitemShut {NoStop}%
\bibitem [{\citenamefont {Baity-Jesi}\ \emph {et~al.}(2017)\citenamefont
  {Baity-Jesi}, \citenamefont {Goodrich}, \citenamefont {Liu}, \citenamefont
  {Nagel},\ and\ \citenamefont {Sethna}}]{BGLNS16}%
  \BibitemOpen
  \bibfield  {author} {\bibinfo {author} {\bibfnamefont {M.}~\bibnamefont
  {Baity-Jesi}}, \bibinfo {author} {\bibfnamefont {C.~P.}\ \bibnamefont
  {Goodrich}}, \bibinfo {author} {\bibfnamefont {A.~J.}\ \bibnamefont {Liu}},
  \bibinfo {author} {\bibfnamefont {S.~R.}\ \bibnamefont {Nagel}}, \ and\
  \bibinfo {author} {\bibfnamefont {J.~P.}\ \bibnamefont {Sethna}},\
  }\href@noop {} {\bibfield  {journal} {\bibinfo  {journal} {Journal of
  Statistical Physics}\ }\textbf {\bibinfo {volume} {167}},\ \bibinfo {pages}
  {735} (\bibinfo {year} {2017})}\BibitemShut {NoStop}%
\bibitem [{\citenamefont {Jin}\ and\ \citenamefont {Yoshino}(2017)}]{JY17}%
  \BibitemOpen
  \bibfield  {author} {\bibinfo {author} {\bibfnamefont {Y.}~\bibnamefont
  {Jin}}\ and\ \bibinfo {author} {\bibfnamefont {H.}~\bibnamefont {Yoshino}},\
  }\href@noop {} {\bibfield  {journal} {\bibinfo  {journal} {Nature
  communications}\ }\textbf {\bibinfo {volume} {8}},\ \bibinfo {pages} {14935}
  (\bibinfo {year} {2017})}\BibitemShut {NoStop}%
\bibitem [{\citenamefont {Jin}\ \emph {et~al.}(2018)\citenamefont {Jin},
  \citenamefont {Urbani}, \citenamefont {Zamponi},\ and\ \citenamefont
  {Yoshino}}]{JUZY18}%
  \BibitemOpen
  \bibfield  {author} {\bibinfo {author} {\bibfnamefont {Y.}~\bibnamefont
  {Jin}}, \bibinfo {author} {\bibfnamefont {P.}~\bibnamefont {Urbani}},
  \bibinfo {author} {\bibfnamefont {F.}~\bibnamefont {Zamponi}}, \ and\
  \bibinfo {author} {\bibfnamefont {H.}~\bibnamefont {Yoshino}},\ }\href@noop
  {} {\bibfield  {journal} {\bibinfo  {journal} {Science Advances}\ }\textbf
  {\bibinfo {volume} {4}},\ \bibinfo {pages} {eaat6387} (\bibinfo {year}
  {2018})}\BibitemShut {NoStop}%
\bibitem [{\citenamefont {Seguin}\ and\ \citenamefont {Dauchot}(2016)}]{SD16}%
  \BibitemOpen
  \bibfield  {author} {\bibinfo {author} {\bibfnamefont {A.}~\bibnamefont
  {Seguin}}\ and\ \bibinfo {author} {\bibfnamefont {O.}~\bibnamefont
  {Dauchot}},\ }\href@noop {} {\bibfield  {journal} {\bibinfo  {journal} {Phys.
  Rev. Lett.}\ }\textbf {\bibinfo {volume} {117}},\ \bibinfo {pages} {228001}
  (\bibinfo {year} {2016})}\BibitemShut {NoStop}%
\bibitem [{\citenamefont {Skoge}\ \emph {et~al.}(2006)\citenamefont {Skoge},
  \citenamefont {Donev}, \citenamefont {Stillinger},\ and\ \citenamefont
  {Torquato}}]{SDST06}%
  \BibitemOpen
  \bibfield  {author} {\bibinfo {author} {\bibfnamefont {M.}~\bibnamefont
  {Skoge}}, \bibinfo {author} {\bibfnamefont {A.}~\bibnamefont {Donev}},
  \bibinfo {author} {\bibfnamefont {F.~H.}\ \bibnamefont {Stillinger}}, \ and\
  \bibinfo {author} {\bibfnamefont {S.}~\bibnamefont {Torquato}},\ }\href
  {\doibase 10.1103/PhysRevE.74.041127} {\bibfield  {journal} {\bibinfo
  {journal} {Physical Review E}\ }\textbf {\bibinfo {volume} {74}},\ \bibinfo
  {eid} {041127} (\bibinfo {year} {2006})}\BibitemShut {NoStop}%
\bibitem [{\citenamefont {van Meel}\ \emph {et~al.}(2009)\citenamefont {van
  Meel}, \citenamefont {Charbonneau}, \citenamefont {Fortini},\ and\
  \citenamefont {Charbonneau}}]{VCFC09}%
  \BibitemOpen
  \bibfield  {author} {\bibinfo {author} {\bibfnamefont {J.~A.}\ \bibnamefont
  {van Meel}}, \bibinfo {author} {\bibfnamefont {B.}~\bibnamefont
  {Charbonneau}}, \bibinfo {author} {\bibfnamefont {A.}~\bibnamefont
  {Fortini}}, \ and\ \bibinfo {author} {\bibfnamefont {P.}~\bibnamefont
  {Charbonneau}},\ }\href@noop {} {\bibfield  {journal} {\bibinfo  {journal}
  {Phys. Rev. E}\ }\textbf {\bibinfo {volume} {80}},\ \bibinfo {pages} {061110}
  (\bibinfo {year} {2009})}\BibitemShut {NoStop}%
\bibitem [{\citenamefont {Kurchan}\ \emph {et~al.}(2012)\citenamefont
  {Kurchan}, \citenamefont {Parisi},\ and\ \citenamefont {Zamponi}}]{KPZ12}%
  \BibitemOpen
  \bibfield  {author} {\bibinfo {author} {\bibfnamefont {J.}~\bibnamefont
  {Kurchan}}, \bibinfo {author} {\bibfnamefont {G.}~\bibnamefont {Parisi}}, \
  and\ \bibinfo {author} {\bibfnamefont {F.}~\bibnamefont {Zamponi}},\
  }\href@noop {} {\bibfield  {journal} {\bibinfo  {journal} {JSTAT}\ }\textbf
  {\bibinfo {volume} {2012}},\ \bibinfo {pages} {P10012} (\bibinfo {year}
  {2012})}\BibitemShut {NoStop}%
\bibitem [{\citenamefont {Maimbourg}\ \emph {et~al.}(2016)\citenamefont
  {Maimbourg}, \citenamefont {Kurchan},\ and\ \citenamefont {Zamponi}}]{MKZ16}%
  \BibitemOpen
  \bibfield  {author} {\bibinfo {author} {\bibfnamefont {T.}~\bibnamefont
  {Maimbourg}}, \bibinfo {author} {\bibfnamefont {J.}~\bibnamefont {Kurchan}},
  \ and\ \bibinfo {author} {\bibfnamefont {F.}~\bibnamefont {Zamponi}},\
  }\href@noop {} {\bibfield  {journal} {\bibinfo  {journal} {Physical review
  letters}\ }\textbf {\bibinfo {volume} {116}},\ \bibinfo {pages} {015902}
  (\bibinfo {year} {2016})}\BibitemShut {NoStop}%
\bibitem [{\citenamefont {Agoritsas}\ \emph {et~al.}(2019)\citenamefont
  {Agoritsas}, \citenamefont {Maimbourg},\ and\ \citenamefont
  {Zamponi}}]{AMZ19}%
  \BibitemOpen
  \bibfield  {author} {\bibinfo {author} {\bibfnamefont {E.}~\bibnamefont
  {Agoritsas}}, \bibinfo {author} {\bibfnamefont {T.}~\bibnamefont
  {Maimbourg}}, \ and\ \bibinfo {author} {\bibfnamefont {F.}~\bibnamefont
  {Zamponi}},\ }\href@noop {} {\bibfield  {journal} {\bibinfo  {journal}
  {arXiv:1903.12572}\ } (\bibinfo {year} {2019})}\BibitemShut {NoStop}%
\bibitem [{\citenamefont {G\"{o}tze}(1999)}]{Go99}%
  \BibitemOpen
  \bibfield  {author} {\bibinfo {author} {\bibfnamefont {W.}~\bibnamefont
  {G\"{o}tze}},\ }\href {http://stacks.iop.org/0953-8984/11/A1} {\bibfield
  {journal} {\bibinfo  {journal} {Journal of Physics: Condensed Matter}\
  }\textbf {\bibinfo {volume} {11}},\ \bibinfo {pages} {A1} (\bibinfo {year}
  {1999})}\BibitemShut {NoStop}%
\bibitem [{\citenamefont {Yoshino}(2012)}]{Yo12}%
  \BibitemOpen
  \bibfield  {author} {\bibinfo {author} {\bibfnamefont {H.}~\bibnamefont
  {Yoshino}},\ }\href@noop {} {\bibfield  {journal} {\bibinfo  {journal} {The
  Journal of Chemical Physics}\ }\textbf {\bibinfo {volume} {136}},\ \bibinfo
  {pages} {214108} (\bibinfo {year} {2012})}\BibitemShut {NoStop}%
\bibitem [{\citenamefont {M\'ezard}\ \emph {et~al.}(1987)\citenamefont
  {M\'ezard}, \citenamefont {Parisi},\ and\ \citenamefont {Virasoro}}]{MPV87}%
  \BibitemOpen
  \bibfield  {author} {\bibinfo {author} {\bibfnamefont {M.}~\bibnamefont
  {M\'ezard}}, \bibinfo {author} {\bibfnamefont {G.}~\bibnamefont {Parisi}}, \
  and\ \bibinfo {author} {\bibfnamefont {M.~A.}\ \bibnamefont {Virasoro}},\
  }\href@noop {} {\emph {\bibinfo {title} {Spin glass theory and beyond}}}\
  (\bibinfo  {publisher} {World Scientific},\ \bibinfo {address} {Singapore},\
  \bibinfo {year} {1987})\BibitemShut {NoStop}%
\bibitem [{\citenamefont {Koumakis}\ \emph {et~al.}(2012)\citenamefont
  {Koumakis}, \citenamefont {Laurati}, \citenamefont {Egelhaaf}, \citenamefont
  {Brady},\ and\ \citenamefont {Petekidis}}]{Kou12}%
  \BibitemOpen
  \bibfield  {author} {\bibinfo {author} {\bibfnamefont {N.}~\bibnamefont
  {Koumakis}}, \bibinfo {author} {\bibfnamefont {M.}~\bibnamefont {Laurati}},
  \bibinfo {author} {\bibfnamefont {S.~U.}\ \bibnamefont {Egelhaaf}}, \bibinfo
  {author} {\bibfnamefont {J.~F.}\ \bibnamefont {Brady}}, \ and\ \bibinfo
  {author} {\bibfnamefont {G.}~\bibnamefont {Petekidis}},\ }\href@noop {}
  {\bibfield  {journal} {\bibinfo  {journal} {Phys.Rev.Lett.}\ }\textbf
  {\bibinfo {volume} {108}},\ \bibinfo {pages} {098303} (\bibinfo {year}
  {2012})}\BibitemShut {NoStop}%
\bibitem [{\citenamefont {Dubey}\ \emph {et~al.}(2016)\citenamefont {Dubey},
  \citenamefont {Hentschel}, \citenamefont {Procaccia},\ and\ \citenamefont
  {Singh}}]{DHPS16}%
  \BibitemOpen
  \bibfield  {author} {\bibinfo {author} {\bibfnamefont {A.~K.}\ \bibnamefont
  {Dubey}}, \bibinfo {author} {\bibfnamefont {H.~G.~E.}\ \bibnamefont
  {Hentschel}}, \bibinfo {author} {\bibfnamefont {I.}~\bibnamefont
  {Procaccia}}, \ and\ \bibinfo {author} {\bibfnamefont {M.}~\bibnamefont
  {Singh}},\ }\href@noop {} {\bibfield  {journal} {\bibinfo  {journal} {Phys.
  Rev. B}\ }\textbf {\bibinfo {volume} {93}},\ \bibinfo {pages} {224204}
  (\bibinfo {year} {2016})}\BibitemShut {NoStop}%
\bibitem [{\citenamefont {Jaiswal}\ \emph {et~al.}(2016)\citenamefont
  {Jaiswal}, \citenamefont {Procaccia}, \citenamefont {Rainone},\ and\
  \citenamefont {Singh}}]{JPRS16}%
  \BibitemOpen
  \bibfield  {author} {\bibinfo {author} {\bibfnamefont {P.~K.}\ \bibnamefont
  {Jaiswal}}, \bibinfo {author} {\bibfnamefont {I.}~\bibnamefont {Procaccia}},
  \bibinfo {author} {\bibfnamefont {C.}~\bibnamefont {Rainone}}, \ and\
  \bibinfo {author} {\bibfnamefont {M.}~\bibnamefont {Singh}},\ }\href@noop {}
  {\bibfield  {journal} {\bibinfo  {journal} {Physical review letters}\
  }\textbf {\bibinfo {volume} {116}},\ \bibinfo {pages} {085501} (\bibinfo
  {year} {2016})}\BibitemShut {NoStop}%
\bibitem [{\citenamefont {Procaccia}\ \emph {et~al.}(2017)\citenamefont
  {Procaccia}, \citenamefont {Rainone},\ and\ \citenamefont {Singh}}]{PRS17}%
  \BibitemOpen
  \bibfield  {author} {\bibinfo {author} {\bibfnamefont {I.}~\bibnamefont
  {Procaccia}}, \bibinfo {author} {\bibfnamefont {C.}~\bibnamefont {Rainone}},
  \ and\ \bibinfo {author} {\bibfnamefont {M.}~\bibnamefont {Singh}},\
  }\href@noop {} {\bibfield  {journal} {\bibinfo  {journal} {Physical Review
  E}\ }\textbf {\bibinfo {volume} {96}},\ \bibinfo {pages} {032907} (\bibinfo
  {year} {2017})}\BibitemShut {NoStop}%
\bibitem [{\citenamefont {Popovi{\'c}}\ \emph {et~al.}(2018)\citenamefont
  {Popovi{\'c}}, \citenamefont {de~Geus},\ and\ \citenamefont {Wyart}}]{PDW18}%
  \BibitemOpen
  \bibfield  {author} {\bibinfo {author} {\bibfnamefont {M.}~\bibnamefont
  {Popovi{\'c}}}, \bibinfo {author} {\bibfnamefont {T.~W.}\ \bibnamefont
  {de~Geus}}, \ and\ \bibinfo {author} {\bibfnamefont {M.}~\bibnamefont
  {Wyart}},\ }\href@noop {} {\bibfield  {journal} {\bibinfo  {journal}
  {Physical Review E}\ }\textbf {\bibinfo {volume} {98}},\ \bibinfo {pages}
  {040901} (\bibinfo {year} {2018})}\BibitemShut {NoStop}%
\bibitem [{\citenamefont {De~Dominicis}\ and\ \citenamefont
  {Kondor}(1983)}]{DK83}%
  \BibitemOpen
  \bibfield  {author} {\bibinfo {author} {\bibfnamefont {C.}~\bibnamefont
  {De~Dominicis}}\ and\ \bibinfo {author} {\bibfnamefont {I.}~\bibnamefont
  {Kondor}},\ }\href@noop {} {\bibfield  {journal} {\bibinfo  {journal}
  {Physical Review B}\ }\textbf {\bibinfo {volume} {27}},\ \bibinfo {pages}
  {606} (\bibinfo {year} {1983})}\BibitemShut {NoStop}%
\bibitem [{\citenamefont {Temesv{\'a}ri}\ \emph {et~al.}(2002)\citenamefont
  {Temesv{\'a}ri}, \citenamefont {De~Dominicis},\ and\ \citenamefont
  {Pimentel}}]{TDP02}%
  \BibitemOpen
  \bibfield  {author} {\bibinfo {author} {\bibfnamefont {T.}~\bibnamefont
  {Temesv{\'a}ri}}, \bibinfo {author} {\bibfnamefont {C.}~\bibnamefont
  {De~Dominicis}}, \ and\ \bibinfo {author} {\bibfnamefont {I.}~\bibnamefont
  {Pimentel}},\ }\href@noop {} {\bibfield  {journal} {\bibinfo  {journal} {The
  European Physical Journal B-Condensed Matter and Complex Systems}\ }\textbf
  {\bibinfo {volume} {25}},\ \bibinfo {pages} {361} (\bibinfo {year}
  {2002})}\BibitemShut {NoStop}%
\bibitem [{\citenamefont {Sommers}(1985)}]{So85}%
  \BibitemOpen
  \bibfield  {author} {\bibinfo {author} {\bibfnamefont {H.-J.}\ \bibnamefont
  {Sommers}},\ }\href@noop {} {\bibfield  {journal} {\bibinfo  {journal}
  {Journal de Physique Lettres}\ }\textbf {\bibinfo {volume} {46}},\ \bibinfo
  {pages} {779} (\bibinfo {year} {1985})}\BibitemShut {NoStop}%
\bibitem [{\citenamefont {Cavagna}\ \emph {et~al.}(1999)\citenamefont
  {Cavagna}, \citenamefont {Garrahan},\ and\ \citenamefont
  {Giardina}}]{Cavagna1999}%
  \BibitemOpen
  \bibfield  {author} {\bibinfo {author} {\bibfnamefont {A.}~\bibnamefont
  {Cavagna}}, \bibinfo {author} {\bibfnamefont {J.~P.}\ \bibnamefont
  {Garrahan}}, \ and\ \bibinfo {author} {\bibfnamefont {I.}~\bibnamefont
  {Giardina}},\ }\href@noop {} {\bibfield  {journal} {\bibinfo  {journal}
  {Journal of Physics A: Mathematical and General}\ }\textbf {\bibinfo {volume}
  {32}},\ \bibinfo {pages} {711} (\bibinfo {year} {1999})}\BibitemShut
  {NoStop}%
\bibitem [{\citenamefont {Franz}\ \emph {et~al.}(2017)\citenamefont {Franz},
  \citenamefont {Parisi}, \citenamefont {Sevelev}, \citenamefont {Urbani},\
  and\ \citenamefont {Zamponi}}]{franz2017}%
  \BibitemOpen
  \bibfield  {author} {\bibinfo {author} {\bibfnamefont {S.}~\bibnamefont
  {Franz}}, \bibinfo {author} {\bibfnamefont {G.}~\bibnamefont {Parisi}},
  \bibinfo {author} {\bibfnamefont {M.}~\bibnamefont {Sevelev}}, \bibinfo
  {author} {\bibfnamefont {P.}~\bibnamefont {Urbani}}, \ and\ \bibinfo {author}
  {\bibfnamefont {F.}~\bibnamefont {Zamponi}},\ }\href@noop {} {\bibfield
  {journal} {\bibinfo  {journal} {SciPost Physics}\ }\textbf {\bibinfo {volume}
  {2}},\ \bibinfo {pages} {019} (\bibinfo {year} {2017})}\BibitemShut {NoStop}%
\bibitem [{\citenamefont {Berthier}\ \emph {et~al.}(2000)\citenamefont
  {Berthier}, \citenamefont {Barrat},\ and\ \citenamefont {Kurchan}}]{BBK00}%
  \BibitemOpen
  \bibfield  {author} {\bibinfo {author} {\bibfnamefont {L.}~\bibnamefont
  {Berthier}}, \bibinfo {author} {\bibfnamefont {J.-L.}\ \bibnamefont
  {Barrat}}, \ and\ \bibinfo {author} {\bibfnamefont {J.}~\bibnamefont
  {Kurchan}},\ }\href@noop {} {\bibfield  {journal} {\bibinfo  {journal}
  {Physical Review E}\ }\textbf {\bibinfo {volume} {61}},\ \bibinfo {pages}
  {5464} (\bibinfo {year} {2000})}\BibitemShut {NoStop}%
\bibitem [{\citenamefont {Folena}\ \emph {et~al.}(2019)\citenamefont {Folena},
  \citenamefont {Franz},\ and\ \citenamefont {Ricci-Tersenghi}}]{FFR19}%
  \BibitemOpen
  \bibfield  {author} {\bibinfo {author} {\bibfnamefont {G.}~\bibnamefont
  {Folena}}, \bibinfo {author} {\bibfnamefont {S.}~\bibnamefont {Franz}}, \
  and\ \bibinfo {author} {\bibfnamefont {F.}~\bibnamefont {Ricci-Tersenghi}},\
  }\href@noop {} {\bibfield  {journal} {\bibinfo  {journal} {arXiv:1903.01421}\
  } (\bibinfo {year} {2019})}\BibitemShut {NoStop}%
\bibitem [{\citenamefont {Lerner}\ \emph {et~al.}(2012)\citenamefont {Lerner},
  \citenamefont {D{\"u}ring},\ and\ \citenamefont {Wyart}}]{LDW12}%
  \BibitemOpen
  \bibfield  {author} {\bibinfo {author} {\bibfnamefont {E.}~\bibnamefont
  {Lerner}}, \bibinfo {author} {\bibfnamefont {G.}~\bibnamefont {D{\"u}ring}},
  \ and\ \bibinfo {author} {\bibfnamefont {M.}~\bibnamefont {Wyart}},\
  }\href@noop {} {\bibfield  {journal} {\bibinfo  {journal} {Europhysics
  Letters}\ }\textbf {\bibinfo {volume} {99}},\ \bibinfo {pages} {58003}
  (\bibinfo {year} {2012})}\BibitemShut {NoStop}%
\end{thebibliography}%

\end{document}